\documentclass[twocolumn]{aastex631}

\hypersetup{linkcolor=black,citecolor=black,filecolor=black,urlcolor=black}
\shorttitle{Star formation in the Trifid Nebula}
\shortauthors{V. M. Kalari} 

\begin{document}

\title{The Young Stellar Population, Distance, and Cloud–Cloud Collision Induced Star Formation Scenario of the Trifid Nebula}

\author{Venu M. Kalari}
\email{venu.kalari@noirlab.edu}
\affiliation{Gemini Observatory/NSF’s NOIRLab, Casilla 603, La Serena, Chile}
\affiliation{Departamento de Astronomia, Universidad de Chile, Casilla 36-D, Santiago, Chile}



\begin{abstract}

The Trifid Nebula is a young, nearby star-forming region where star formation is proposed to have been triggered by cloud-cloud collision (CCC), based on observations of molecular clouds. It offers a unique opportunity to test whether the CCC hypothesis is supported by the spatial distribution and star formation chronology of young stars.  

We present the first study of the optically visible pre-main sequence (PMS) population of the region using {\it ri}H$\alpha$ imaging and {\it Gaia} astrometry. Combined with an analysis of young stellar objects (YSOs) using infrared imaging, we capture the spatial distribution and star formation chronology of the young stellar population. From the analysis, 15 Flat/Class\,I YSOs, 46 Class\,II YSOs, and 41 accreting PMS stars are identified (diskless/non-accreting sources are not included in the analysis). The distance based on {\it Gaia} parallaxes is $\sim$1250\,pc, significantly closer than previously reported. The Class\,II YSOs and PMS stars ($\sim$1.5\,Myr old) are spread toward the edge of the molecular clouds. They are slightly younger than the estimated crossing time of $\sim$2.7\,Myr and closer to the estimated dynamical age $\sim$0.85\,Myr. Younger Class\,I\,YSOs are more concentrated spatially. There exists a cavity devoid of young stars where the two clouds overlap. This evidence suggests that the current generation of stars formed after the collision of two clouds $\sim$1\,Myr ago, and this result can be corroborated using future spectroscopic studies.

\end{abstract}



\section{Introduction}

\begin{figure*}
\center 
\plotone{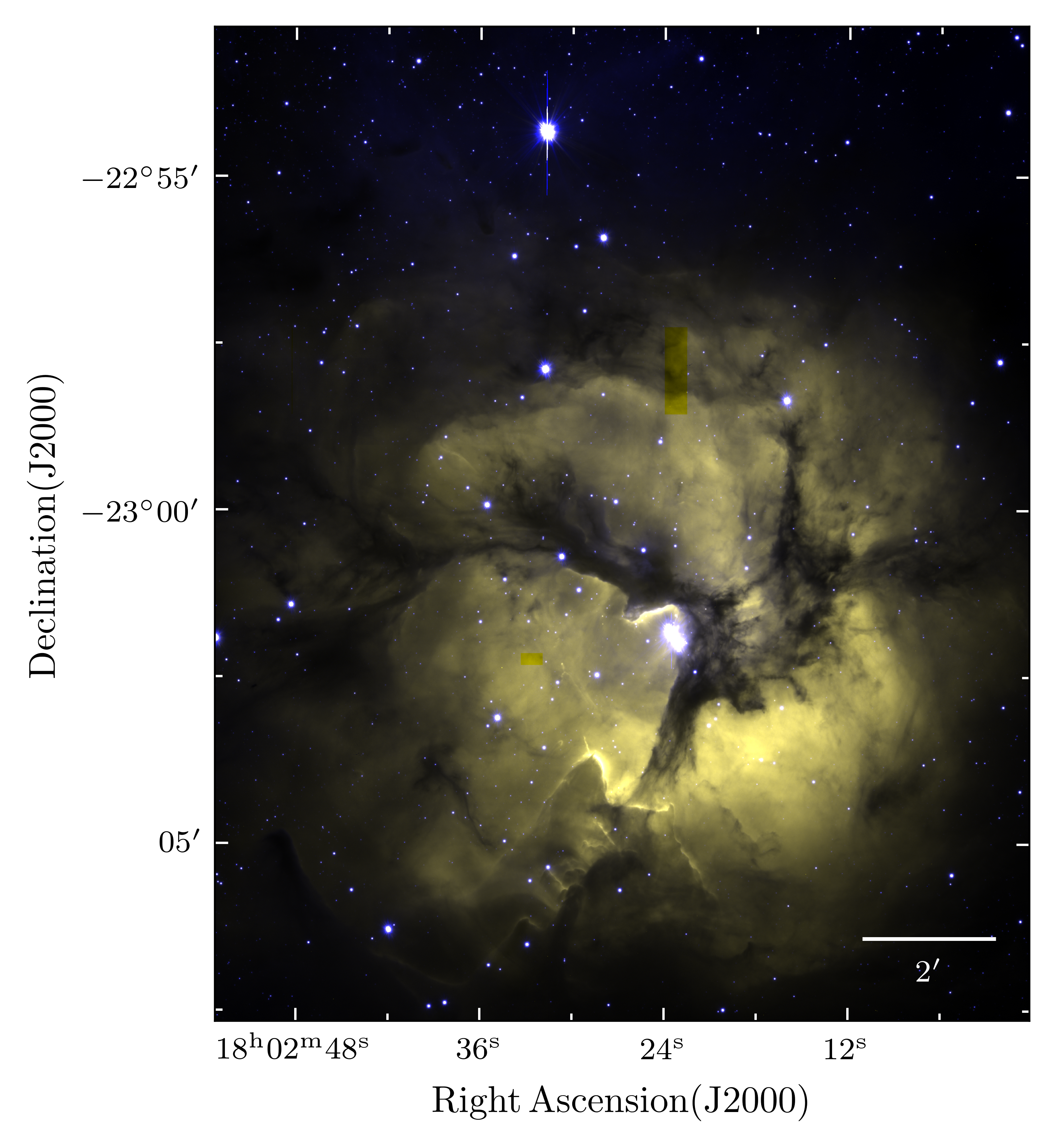}
\center
\caption{The Trifid nebula as an {\it rgb} image from VPHAS+ $i$, H$\alpha$, and $r$ images, respectively. The area shown is centred on right ascension (J2000) 18$^{h}$02$^{m}$26$^{s}$8 and declination (J2000) $-$23$\degr$00$\arcmin$12$\arcsec$ with a field of view 12$\arcmin$2\,$\times$\,14$\arcmin$9, and covers the area studied in this work. North is up and east to the left. The central ionising O7.5V star in the HD\,164492 system is visible as the bright central star. Nebulosity, varying on small spatial scales is also seen. }
\label{image}
\end{figure*}

There remain many open questions in the field of star formation. In the current paradigm, star formation is a dynamical process, occurring primarily due to gravitational instabilities seeded by turbulence in molecular clouds \citep{mckee}. A spectacular trigger for gravitational instabilities and the eventual collapse of molecular clouds into stars is cloud-cloud collision (CCC). It is has been argued that the collision of one or more molecular clouds can potentially cause over-densities, leading to gravitational collapse and subsequent star formation \citep{loren1976}. Recent observations at sub-mm wavelengths have presented evidence that CCC is a viable mechanism to form stars and young clusters across the Milky Way and in external galaxies \citep{fukui}. 
 
The young star-forming region Trifid Nebula (or M\,20, NGC\,6514) is a proposed site of CCC induced star formation \citep{torii11, torii17}, based on the systemic velocities and masses of two molecular clouds in the region. These two clouds are thought to have collided $\sim$1\,Myr ago, producing the young stellar population observed in the region. M\,20 is one of the youngest sites ($\sim$0.3--1\,Myr; \citealt{rho08}) proposed to have formed via CCC. It is thought to lie between 1.6--2.7\,kpc away \citep{rho08, cambresy11}, either in the Sagittarius or Scutum arm. This combination of proximity and age allows one to observe and characterise the young low-mass stellar population of the region, a possibility not afforded in studies of more distant sites of CCC \citep{fukui}. It is therefore an ideal avenue to gauge whether the young stellar population is supportive of an induced star-formation scenario.   

The stellar population of the region has not been studied in the optical. An O7.5V star, HD\,164492A (with a spectroscopic age of 0.6\,Myr; \citealt{164492Age}) ionises the optically observed nebula (see Fig.\,1), and lies south west of the crux of the three bisecting dust lanes. The extinction law in the region is anomalous at $R_V=5.5$ \citep{cambresy11}. The combination of nebulousity, extinction, and a line-of-sight towards the Galactic plane have hindered optical observations. 
Although there have been no optical studies of the young stellar population in the region, multiple studies at radio, sub-mm, mid-infrared (mIR), and near-infrared (nIR) have been conducted \citep{yusefzadeh00, lefloch, rho01, rho06, rho08, mystix}. These studies have revealed a rich young stellar population at a very nascent stage of star formation ($\lesssim$1\,Myr old); a population of even younger star-forming cores \citep{tapia}, along with evidence for photo-ionisation affecting on-going star formation \citep{yusefzadeh05}  

The different wavelengths trace the varying stages of on-going star formation in the region; from the molecular clouds visible at sub-mm and radio wavelengths, to the young protostars in the infrared (IR). Studies of the optically visible pre-main sequence (PMS) stars and the stellar cluster would complete the circle, helping to piece together the star formation picture. With a combination of deep optical photometry, and astrometry from the {\it Gaia} space telescope, it is now possible to both observe the low-mass optically visible PMS stars and disentangle them from line-of-sight of contaminants. Thus, when combined with archival IR data, we are now armed with observations to characterise the young stellar population of the Trifid Nebula, and examine how they relate to the proposed CCC triggered star formation scenario.

In this paper we identify and characterise the young stellar population in the Trifid Nebula using IR and optical observations, combined with {\it Gaia} astrometry to investigate whether the triggered star formation scenario proposed in the literature can explain the observed spatial distribution and star formation chronology of the young stellar population. This paper is organised thus; in Section 2, a description of the data used in this work is presented. Sections 3 and 4 describe the identification and classification of the IR young stellar population, and optically visible PMS stars respectively. In Section 5 a discussion on the star-formation scenario in the Trifid nebula is presented.

\section{Data}

\begin{figure}
\center 
\includegraphics[width=8cm]{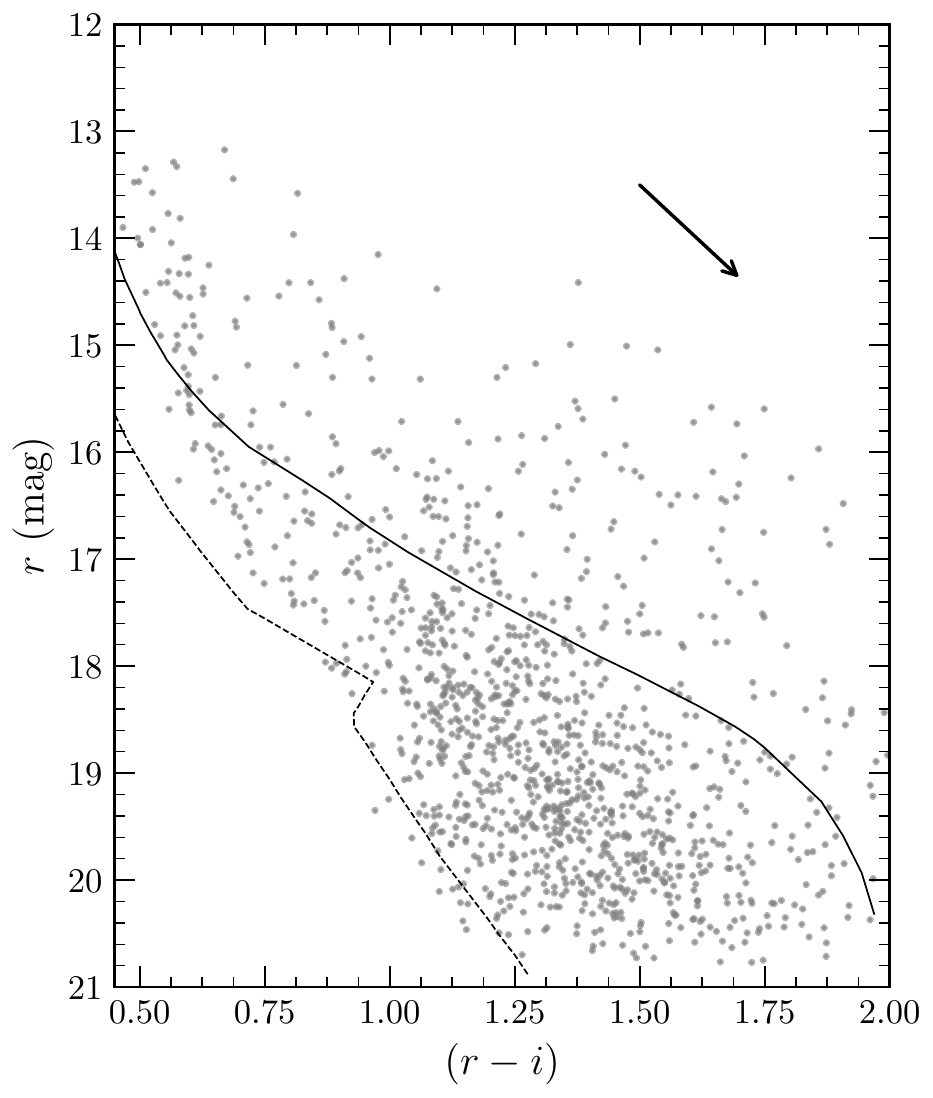}
\center
\caption{Observed ($r-i$) vs. $r$ colour magnitude diagram of all sources studied in this work. A reddening vector of $A_V$=1\,mag, $R_V$=5.5, is shown in the top right. \cite{bress12} isochrones of 1 (solid) and 10\,Myr (dashed) at a distance of 1250\,pc and $A_V$\,=\,1.3\,mag are shown. }
\label{cmdyes}
\end{figure} 
\subsection{Area of study}


To identify and characterise the young stellar population in the star-forming region M\,20, we define our area of study as all sources falling within a bounding box of 12$\arcmin$2\,$\times$\,14$\arcmin$9 region centred on right ascension 18$^{h}$02$^{m}$26$^{s}$8 and declination $-$23$\degr$00$\arcmin$12$\arcsec$ (see Fig. 1). The area definition is slightly smaller than previous studies at nIR or X-ray wavelengths \citep{rho01,rho06, mystix}, and completely includes only the central Trifid nebula, and the blister region located immediately above it. Because M\,20 lies directly in our line of sight towards the Galactic plane, multiple populations in further arms will be confused with optical members. The optical nebulosity as seen in M\,20 can prevent the detection of more distant contaminants in the plane, providing an upper blue envelope in the colour-magnitude diagram. In addition, the area chosen adequately covers the region where CCC and subsequent star formation are thought to have occurred \citep{torii11}. Choosing a slightly larger area of study while potentially enabling a more detailed understanding of the extent of the region, will include more distant non-members which cannot be separated easily, justifying our choice of boundaries. 

\subsection{VPHAS+ {\it ugri}H$\alpha$ imaging}

The VPHAS+ (VST/Omegacam Photometric H$\alpha$ Survey of the Southern galactic plane and bulge) survey observed the Trifid Nebula in {\it ugri}H$\alpha$ filters, and is comprehensively outlined in the survey paper by \cite{drew14}. The data used in this publication is part of the VPHAS+ third data release. The reader is referred to the release document{\footnote{http://www.eso.org/rm/api/v1/public/releaseDescriptions/106}} for details of the data reduction procedure. 
A {\it rgb} image created using the final reduced images is shown in Fig.\ref{image} to illustrate the quality of the imaging and the area studied in this work. 
Clearly visible are the three central dust lanes, and the photoionized regions towards the south. 

Aperture photometry was performed on these reduced images using a dedicated pipeline run by the Cambridge Astronomy Survey Unit. Attention is drawn to the dedicated software {\it nebuliser} used in the photometric pipeline to track variable background emission on a sliding scale around 15$\arcsec$. Objects that suffer from improper background subtraction due to rapidly varying nebulousity on smaller scales are flagged using a curve-of-growth analysis as possibly non-stellar. The final catalogue available from the archive is calibrated in the AB magnitude system. A global Vega magnitude calibration is achieved across all filters by cross matching sources overlapping with the VPHAS+ DR2 data
, which was globally calibrated on the Vega magnitude scale. The global shifts are 1.084, $-$0.055, 0.138, 0.461, and 0.316 in {\it ugri}H$\alpha$ filters respectively. For the remainder of the study, the magnitudes reported are in the Vega magnitude scale.

From the publicly available catalogue, we select all sources falling within the area of interest. We apply further selection criteria of (i) $r>$13 to remove saturated sources, (ii) photometric uncertainties in $ri$ less than 0.1\,mag, and in H$\alpha$ less than 0.15\,mag to keep low propagated errors, (iii) sources classified as stellar, or star-like in broadband $ri$ photometry to remove extended objects. For H$\alpha$ photometry, mild degradation of the point-spread function may lead to spurious classifications as extended objects. To avoid this and not reject acceptable H$\alpha$ measurements, the classification in H$\alpha$ is relaxed to extended sources. Instead, a further visual check using unsharp masking on selected sources is imposed to eliminate any sources affected by sharply varying nebulousity (see \citealt{Kalari15}), (iv) sources with average photometric confidence greater than 90 to remove sources falling in the chip gaps. Note that we do not impose any quality criteria on $ug$ photometry. The cleaned sample contains 9\,227 unique sources with $ri$H$\alpha$ photometry meeting the quality criteria. 
The colour magnitude diagram generated using the data is shown in Fig.\,\ref{cmdyes}. The saturation limit of VPHAS+ photometry means that the bright, early type members of the Trifid nebula are missing from our analysis. Unlike other nearby star-forming regions near the Galactic plane, there is no upper blue envelope due to the nebulosity. More distant Galactic contaminants are visible, particularly at the edges of the nebula.

Finally, a result of our selection criteria is that stars near the nebulous centre, and along the dust lanes are preferentially removed, as they lie along regions of very rapidly spatially varying nebulousity. Optical photometry from the {\it Gaia} telescope also shows similar spatial density to the VPHAS+ observations, suggesting that observations at the optical wavelength towards the central region suffer greatly from the spatial varying nebulousity. The consequences of this spatial density variation in the optical on our results is discussed in Section\,4.6.

\subsection{{\it Gaia} EDR\,3 astrometry}
Astrometric data from the {\it Gaia} early data release 3 (EDR3; \citealt{gaia3}) are cross-matched to the VPHAS+ data. {\it Gaia}\,EDR3 data contain parallaxes ($\pi$), proper motions vectors in right ascension ($\mu_{\alpha}$cos$\delta$, hereafter referred to as $\mu_{\alpha}$), declination ($\mu_{\delta}$), and the associated photometry in custom filters measured from the first three years of {\it Gaia} observations (2014-2017) in the 2016 reference epoch. 
{\it Gaia} EDR3 photometry has a detection threshold of $G\sim$\,21\,mag, with the very bright ($G<$\,7\,mag) and high proper-motion ($>$\,0.6 arcsec\,yr$^{-1}$) stars incomplete. The detection limit of {\it Gaia} corresponds approximately to the 3$\sigma$ threshold of the VPHAS+ photometry in the $ri$ filters. Errors in EDR3 data release on the proper motions are believed to be around $\pm$0.02--1.4\,mas depending on source magnitude, and on parallax around $\pm$0.02-1.3\,mas. A comparison with extragalactic sources indicates an overall negative bias of $-$17\,$\mu$as \citep{gaia3} on the parallax, which was added to our parallax measurements. 

To perform cross-matching with the VPHAS+ sample, a minimum cross-match radius of 0.1$\arcsec$ was set considering the {\it Gaia} positional accuracy (0.059$\arcsec$), and separation between the different epochs (also 0.059$\arcsec$). Sources were initially matched using a maximum radius of 2$\arcsec$ (with both catalogues in the J2000 epoch), where the mean separation was found to be 0.077$\arcsec$. To determine the incidence of spurious sources, the radius was incrementally reduced by 0.05$\arcsec$, while also counting the number of stars with magnitude differences between the {\it Gaia}\,$G$\,band, and the VPHAS+\,$i$ magnitude (applying an offset of 0.6\,mag to bring approximately to a relation of unity) was greater than 1\,mag (accounting for variability; note that $r$-band is not used as it includes the H$\alpha$ line). A best-match radius of 0.3$\arcsec$ is chosen as 98\% of matches within 1$\arcsec$ are recovered, but no sources with large magnitude offsets are picked up. The renormalised unit weight error (RUWE) cut of RUWE$<1.4$ was applied to remove sources with poor astrometry. As a result, 8\,965 sources with high-fidelity astrometry and photometry from {\it Gaia} EDR3 that have cross-matches in VPHAS+ comprise our dataset. This is the dataset that will be utilised for identifying and characterising the young stellar population visible at optical wavelengths. 

\subsection{nIR and mIR imaging}


\begin{figure*}
\center 
\includegraphics[width=16cm]{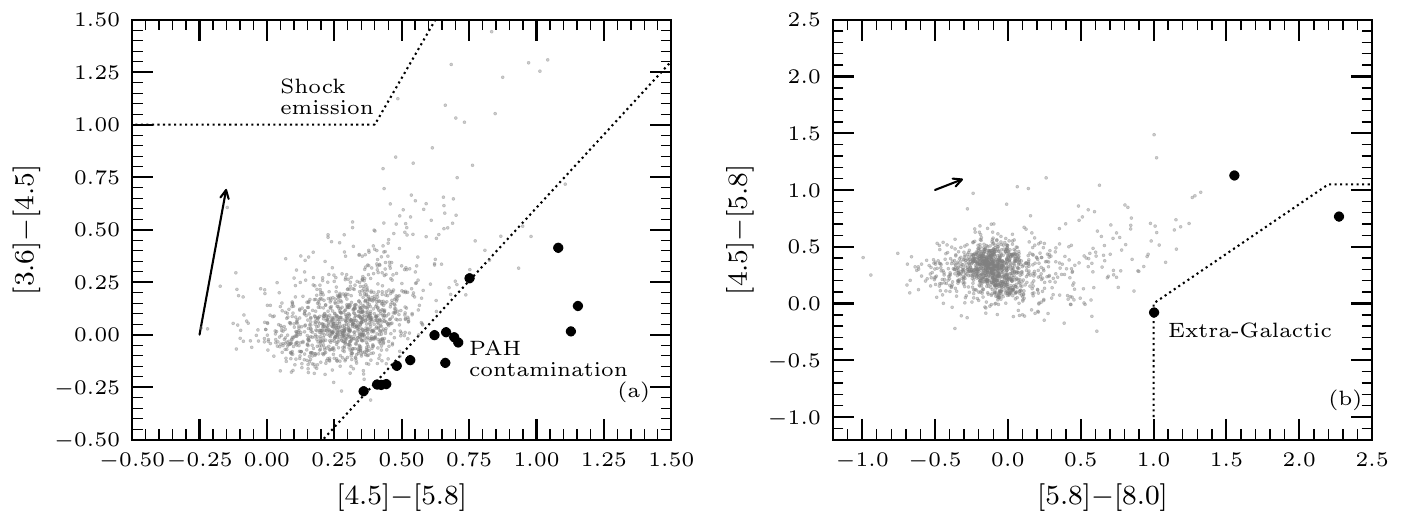}
\center
    \caption{($a$): [4.5]$-$[5.8] vs. [3.6]$-$[4.5] colour-colour diagram of all mIR sources are shown as circles. The location of sources affected by PAH contamination and shock emission are shown following \cite{gut}. Stars removed following the criteria outlined in the text as contaminated photometry are marked by filled circles. ($b$) [5.8]$-$[8.0] vs. [4.5]$-$[5.8] colour-colour diagram of all sources are shown. Sources that are probable external galaxies following \cite{gut} classification criteria are marked by filled circles. The reddening vector for $A_K$=5\,mag is also shown. The mid-infrared reddening law of \cite{inde05} was used.}
\label{iracremove}
\end{figure*} 

The nIR sample consists of all sources with $JHK$s photometry from the UKIDSS (UKIRT Infrared Deep Sky Survey; \citealt{ukidss}) survey. To this sample, we discarded sources with non-stellar profiles (akin to the classification schema in the VPHAS+ photometry), or lacking photometry in any of the three bands, but applied no further selection criteria. 
This catalogue is cross-matched with the VPHAS+--{\it Gaia} sample. Given the detection limit of UKIDSS ($\sim$19.7 mag in $J$), the optical sample misses some of the lower mass mid-M spectral type stars (assuming average extinction, with no impact of nebulosity) that are captured by the nIR photometry. However, nebulosity towards the core of the Trifid nebula greatly impacts our detection of sources in the region. The effect of this incompleteness in optical photometry compared to the nIR on our final results is discussed in further detail in Section\,4.6.

To capture the disc evolutionary stages using data from the mIR, we utilise archival photometry from the GLIMPSE (The Galactic Legacy Infrared Midplane Survey Extraordinaire; \citealt{glimpse}) survey catalogue. GLIMPSE details photometry in the [3.6], [4.5], [5.8] and [8.0] IRAC (InfraRed Array Camera) filters along the Galactic plane, taken using the {\it Spitzer} space telescope. Only sources with no close sources within 2$\arcsec$ and with no photometric quality issue are considered. No photometric uncertainty limit was applied to the data set. To this sample, we performed a cross-match search against the MIPSGAL survey (MIPS Galactic Plane Survey; \citealt{atlasgal}) to find any sources having photometry at 24$\mu$m, which was taken using the MIPS (Multiband Imaging Photometer for {\it Spitzer}) instrument. Note that both the mIR catalogues already contain cross-matches to the 2MASS $JHK$s \citep{cutri03} photometry catalogue. We cross-match the mIR catalogue with the UKIDSS, and VPHAS+--{\it Gaia} datasets using a radius of 2$\arcsec$ (the mean effective resolution of the mid-infrared photometry), and performed a sanity check on the cross-matches by removing any sources which have GLIMPSE provided $K$s band photometry deviant by more than 0.5\,mag from the UKIDSS survey $K$s photometry.

\section{Young stellar objects}

Young stellar objects (YSOs) may be discriminated from the stellar population on the basis of their IR properties. Depending on their evolutionary stage, YSOs are likely to be surrounded by either circumstellar discs, or envelopes which absorb and re-radiate stellar luminosity in IR portion of the electromagnetic spectrum. At these wavelengths, the colours of YSOs, and the shape of their spectral energy distribution (SED) significantly differ with respect to the vast majority of observable stars. The database of nIR and mIR photometry spanning the wavelength range of 1.25$\mu$m--24$\mu$m described in Section\,2.4 is used to both identify YSOs, and classify their evolutionary stage in this section.



\subsection{Identification of young stellar objects}

To identify probable YSOs, we select from the {\it Spitzer} IRAC sample only sources with photometry in all four bands. In addition, nIR and 24$\mu$m photometry of these sources is also used for the analysis when available. 
The main source of contamination when attempting to identify YSOs against the Galactic plane is dealing with foreground/background stars, and PolyAromatic Hydrocarbon (PAH) and background galaxies. Most foreground/background stars do not have IR excesses, and are centered around zero in {\it Spitzer} colour-colour planes \citep{megeath}. Removal of AGB (Asymptotic Giant Branch) stars is less straightforward, as they do exhibit excesses similar to YSOs in some IR colours. Similarly, background galaxies have colours falling in the region occupied by YSOs in the mIR colour-colour plane. Active star-forming galaxies in particular have colours that are dominated by PAH emission, and they are bright in the [8.0] filter (due to the 7.7--8.2$\mu$m PAH feature). AGN (active galactic nuclei), are similarly bright at 8$\mu$m and also overlap YSOs in the colour-colour plane. In addition, the photometry also may be affected by unresolved knots caused due to shock emission from YSOs, or from PAH emission contaminating the apertures of sources. The latter two represent significant sources of photometric contamination in star-forming regions. To remove extra-galactic sources, and contaminated photometry, we adopt the criteria outlined in \cite{gut}, appendix A. Note that a reddening correction corresponding to an absolute extinction, $A_V$ of 1.3\,mag \citep{rho08}, following the reddening law ($R_V$) of 5.5 \citep{cambresy11} was applied. The results of this exercise are shown in Fig.\,\ref{iracremove}. Overall, $\sim$20 sources were removed following this criteria. In addition, to identify stellar contaminants, a cutoff of [4.5]$-$[8.0]$<$0.5, and [8.0]$<5$\,mag following \cite{harvey} was applied. The cutoff in magnitude is scaled to a distance of 1250\,pc (see Section\,4.3 for details on the chosen distance). These sources are not removed from the sample immediately, but this cut-off is applied when classifying YSOs.

\subsection{Classification of young stellar objects}

\begin{figure*}
\center 
\includegraphics[width=16cm]{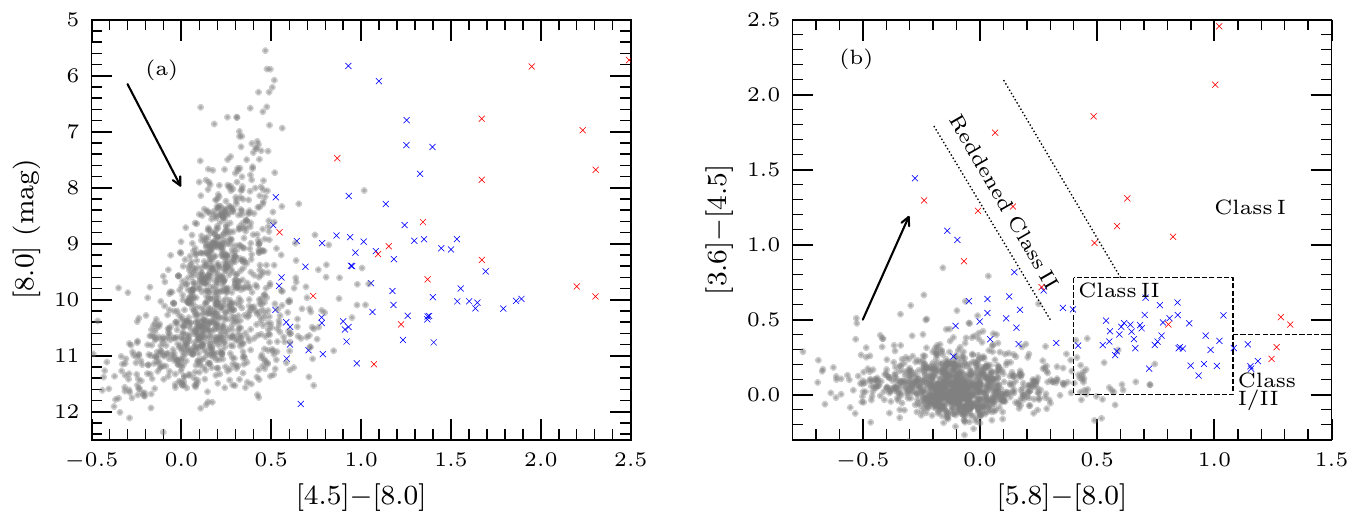}
\center
\caption{($a$): [4.5]$-$[8.0] vs. [8.0] colour-magnitude diagram of mIR sources are shown as circles. Sources identified as Class I and Class II YSOs are marked as red and blue crosses respectively. ($b$) [5.8]$-$[8.0] vs. [3.6]$-$[4.5] colour-colour diagram of all sources. The location of Class II and Class I sources following \cite{allen} and \cite{megeath} classification criteria are shown. The reddening vector for $A_K$=5\,mag is also shown, based on the \cite{inde05} reddening law.   }
\label{iraclcassify}
\end{figure*} 

There exist multiple criteria outlined in the literature to identify and classify YSOs using either the value of the IR SED slope ($\alpha$), or positions of sources in colour-colour diagrams. These methods are similar in philosophy, but differ in details. A full review and examination of these criteria is beyond the scope of this paper, and we base the analysis in this paper on a combination of the colour, and slope classification criteria outlined in \cite{megeath} and \cite{allen}, and \cite{harvey} respectively. Firstly, we classify all sources that were not identified as extra-galactic sources, or having photometry contaminated by PAH or shock emission using the value of $\alpha$, defined as
    \begin{equation}
            \alpha = \frac{d\,log\lambda F_{\lambda}}{d\,log(\lambda)}.
\end{equation}
Here $\lambda$ is wavelength and $\lambda F_{\lambda}$ is flux density in wavelength. To determine $\alpha$, we fit the fluxes of all sources longward of 2$\mu$m. For all sources, we require at least photometry in all IRAC bands. The photometry was dereddened assuming the mean extinction towards the region. The slope was calculated using a weighted least squares fit to the available photometry. Here, the weight applied was the inverse of the squared photometric errors. The fit error from a standard weighted least squares fit is also calculated. 

Following the classification scheme of \cite{harvey}, we classify sources as YSOs based on their $\alpha$ value. The YSO classes are;

\begin{enumerate}
  \setlength{\itemsep}{0pt}
  \setlength{\parskip}{0pt}
\item[] Class\,I: $\alpha>$ 0.3 
\item[] Flat: $0.3\geq\alpha\geq-$0.3 
\item[] Class\,II: $-0.3>\alpha\geq-$1.6 
\end{enumerate}
Note that we do not attempt to classify Class\,III sources. Class\,III sources (YSOs with no or transition discs) cannot be differentiated from the bulk of the stellar population as they lack IR excesses. While in the literature Class\,III sources have been classified as non-contaminants with $\alpha<-$1.6, we refrain from doing so here as the Trifid nebula lies in a direct sight line towards the Galactic plane, and even in mIR is heavily contaminated by background/foreground field stars \citep{mystix}. Therefore, classifying Class\,III sources based on $\alpha$ would lead to spurious detections.  


{\it Spitzer} images of the sources were visually inspected to detect any that may be affected by extended emission, nearby contaminating bright sources (recall that we discard previously any sources having counterparts within 2$\arcsec$, the mean full-width half maximum of the photometry), or are elongated. Three sources that are extended, or have a nearby bright source within 2$\arcsec$5 are removed. Note that the bulk of our sample has $\alpha\sim-2.5$ to $-3$, agreeing with the expectation for main-sequence stars. 

\begin{deluxetable*}{llllllllcc}
\tablenum{1}
\tablecaption{Infrared photometry of young stellar objects in the Trifid Nebula. \label{tab:messier}}
\tablewidth{0pt}
\tablehead{
\colhead{GLIMPSE ID} & \colhead{$J$} & \colhead{$H$} & \colhead{$K$s} &
\colhead{[3.6]} & \colhead{[4.5]} & \colhead{[5.8]} &
\colhead{[8.0]} & \colhead{$\alpha$} & \colhead{Class} \\
\colhead{} & \colhead{(mag)} & \colhead{(mag)} & \colhead{(mag)} & \colhead{(mag)} & \colhead{(mag)} & \colhead{(mag)} & \colhead{(mag)} & \colhead{} & \colhead{}
}
\startdata
  G006.9115-00.2629 &  &  &  & 12.07$\pm$0.06 & 10.30$\pm$0.04 & 9.26$\pm$0.04 & 9.18$\pm$0.09 & 1.02$\pm$0.3 & I\\
    G006.9258-00.2629 & 13.27$\pm$0.02 & 12.33$\pm$0.02 & 11.73$\pm$0.033 & 10.88$\pm$0.05 & 10.55$\pm$0.04 & 10.17$\pm$0.06 & 9.08$\pm$0.06 & $-$0.9$\pm$0.2 & II\\
  G006.9292-00.2460 & 16.30$\pm$0.10 & 14.08$\pm$0.05 & 12.13$\pm$0.02 & 9.53$\pm$0.03 & 8.46$\pm$0.03 & 7.60$\pm$0.03 & 6.76$\pm$0.02 & 0.2$\pm$0.1 & Flat\\
  G006.9451-00.3318 &  & 14.18$\pm$0.07 & 12.45$\pm$0.05 & 10.60$\pm$0.17 & 10.00$\pm$0.13 & 9.32$\pm$0.04 & 8.96$\pm$0.03 & $-$1.49$\pm$0.3 & II\\
  G006.9637-00.3104 & 11.15$\pm$0.02 & 10.75$\pm$0.02 & 10.42$\pm$0.02 & 9.7$\pm$0.08 & 9.23$\pm$0.04 & 8.26$\pm$0.082 & 6.97$\pm$0.18 & 0.27$\pm$0.4 & Flat\\
  \enddata
\tablecomments{Five selected rows are shown to display the form and contents of the table. The full table is only available electronically, along with the source coordinates. The 24$\mu$m magnitude for G007.0317-00.2845 is 
1.7$\pm$0.03, G006.9221-00.2513 is 2.29$\pm$0.02, G007.0097-00.2542 is 2.87$\pm$0.18, G006.9292-00.2460 is 4.14$\pm$0.03, G007.0574-00.2701 is 4.79$\pm$0.12 mag respectively.}
\end{deluxetable*}

After the SED slope analysis, an additional colour selection is applied to cut any remaining contaminants (e.g., AGB stars, reddened main-sequence stars). The criteria outlined in \cite{megeath} and \cite{allen} is utilised, and the results are shown in Fig.\,\ref{iraclcassify}. Here, the approximate colours of Class\,II, Class\,I, a combination of Class\,I/II sources, and reddened Class\,II YSOs given by \cite{megeath} and \cite{allen} are shown. The location of sources classified based on $\alpha$ are also shown. The loci of main sequence sources without excess in the IR are centred around zero as expected, with a spread due to reddening differences and CO absorbing giants. The positions of Class\,II sources are demarcated by the bounding box, which is reproduced by models of discs with varying accretion rates, and inclinations around young and low-mass stars \citep{d05}. The empirical boundary between Class\,II and Class\,I sources is also given, and their colours are of protoestellar objects with infalling envelopes. The location of Class\,II sources affected by [8.0] excess emission is marked in the region of Class I/Class II sources, as is the location of Class\,II sources affected by extreme reddening. While in general, the location of these sources agrees with the expectation based on their colour criteria, we note that 17 sources fall blueward of the region demarcated as either Class\,II or reddened Class\,II sources. Sources falling in this location are spatially preferentially located towards the edges of the nebula, directly in the sight of the Galactic plane, and have colours [4.5]$-$[8.0]$<$1 \citep{harvey}, similar to AGB stars. These are likely reddened AGB stars in the Galactic plane. To remove them, these 17 sources falling outside of the expected colour boundaries for YSOs from \cite{megeath} and \cite{allen} are removed. 

\subsection{Final sample of YSOs}

Our final sample consists of 46 Class\,II YSOs, 9 Flat, and 6 Class\,I YSOs. The complete sample along with archival photometry is given in Table\,1. A source of contamination in our final sample is background/foreground AGB stars, as sources resembling colours of galaxies, or contaminated aperture photometry were removed, and the images of the final classified sources were visually inspected. The contamination rate in our sample may be judged by adopting the locations of AGB stars in colour-magnitude diagrams as found by \cite{harvey}. In that paper, it is suggested that some AGB stars which may have similar SED slopes to YSOs have colours [4.5]$-$[8.0]$<1$. These AGB stars represent the largest source of contamination in our sample, given the sight line towards the Galaxy. 12 YSOs have colours bluer that AGB stars (but redder than [4.5]$-$[8.0]$>$0.5), and these represent $\sim$20\% of our sample. This represents an upper limit on the estimate of the contamination in the sample, given that otherwise the sources meet the SED and colour classification criteria.

The completeness of the sample is difficult to judge straightforwardly, as there are various factors affecting it. In particular, the saturation of photometry towards the centre of the Trifid, and the large-scale differences in the nebulosity. These cannot be straightforwardly estimated even using a control field given the location near the Galactic bulge. A similar conclusion was arrived at by the study of \cite{mystix}, who identified YSOs in the Trifid nebula based on a combination of modelling IR SEDs, and X-ray emission. 

As an alternative, we estimate the recovery rate of our sample against the one selected by \cite{mystix}. They identified 41 Stage II/III, and 22 Stage I YSOs having {\it Spitzer} photometry in all IRAC bands within the spatial boundary used in this work by modelling the available photometry against the YSO models of \cite{rob}. From their sample, we find 28 Stage II/III, and 11 Stage I YSOs have cross matches in our final YSO sample. In their remaining Stage II/III sample, we find that two have mIR colours resembling PAH contaminated sources, while two are likely stellar contaminants ([4.5]-[8.0]$<$0.5) according to our analysis, and three had $\alpha$ values of Class\,II YSOs, but were removed from the final sample based on their positions in the colour-colour plane resembling contaminants (see Section\,3.2). The remaining six had $\alpha$ values between $-1.7$ and $-2.1$ (two of which fell in the contaminant region in the colour-colour plane). This distribution is expected given that the Stage and SED slope classifications are not analogous, and depend on various factors \citep{rob}. In addition, the Stage classification assumes that sources with $\alpha>-2$ are Stage II, and $\alpha<-2$ are Stage III. Out of the 11 Stage I sample from \cite{mystix} not having counterparts in this sample, 1 star was not present in our source catalogue, while 7 were discarded as PAH or shock emission contaminants. The three remaining sources have colours resembling reddened AGB stars and were removed from our final sample. Therefore, our sample compares well to the YSO catalogue of \cite{mystix} once the constraints chosen in this work are applied.

Given the contamination rate, and absence of a $\chi^2$ fitting value (to adjudicate the quality of the classification) for the sources in \cite{mystix}, archival infrared Class\,III sources from that catalogue are not included here. Similarly, X-ray bright Class\,III sources from that survey were not included as the X-ray detection threshold chosen is considerably faint, and a significant fraction ($\sim$35\%) of Class\,III X-ray sources do not have counterparts at other wavelengths, suggesting they could be faint background sources. In addition, a distance value of 2700\,pc was (see Section 4.3) used to calculate membership probabilities and absorption, possibly leading to the inclusion of more distant X-ray sources.

\section{Pre-main sequence stars}

As YSOs evolve, their surrounding material collapses into circumstellar discs to conserve angular momentum. Accretion of mass from the disc towards the central stellar source via the stellar magnetosphere is necessary to gain enough material to initiate fusion. This accretion process results in a combination of unique signatures including excess ultraviolet continuum, and line emission, notably in H$\alpha$ \citep{gullbring98}. It is through this excess H$\alpha$ emission that accreting young PMS stars can be readily identified. 

\begin{figure}
\center 
\includegraphics[width=8cm]{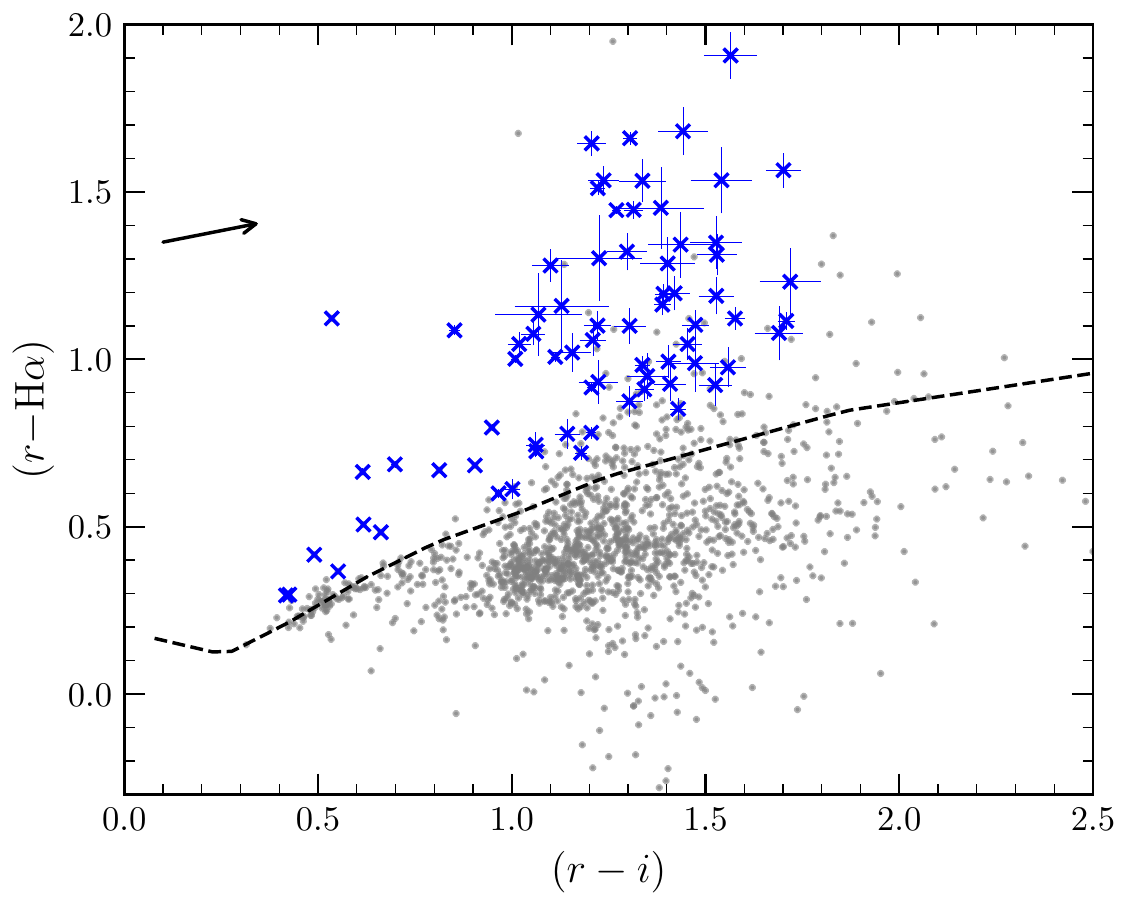}
\center
\caption{($r-i$) vs. ($r-$H$\alpha$) colour-colour diagram of all sources considered in the work, shown as dots. A reddening vector of 1\,mag is shown in the top left. The dashed line represents the main-sequence colour locus calculated from the \cite{pick98} synthetic spectra, reddened by $A_V=1.3$\,mag. Stars selected as accretors based on the {\rm EW}$_{\rm{H}\alpha}$ criterion are marked by blue crosses.}
\label{riha}
\end{figure} 

\begin{figure}
\center 
\includegraphics[width=8cm]{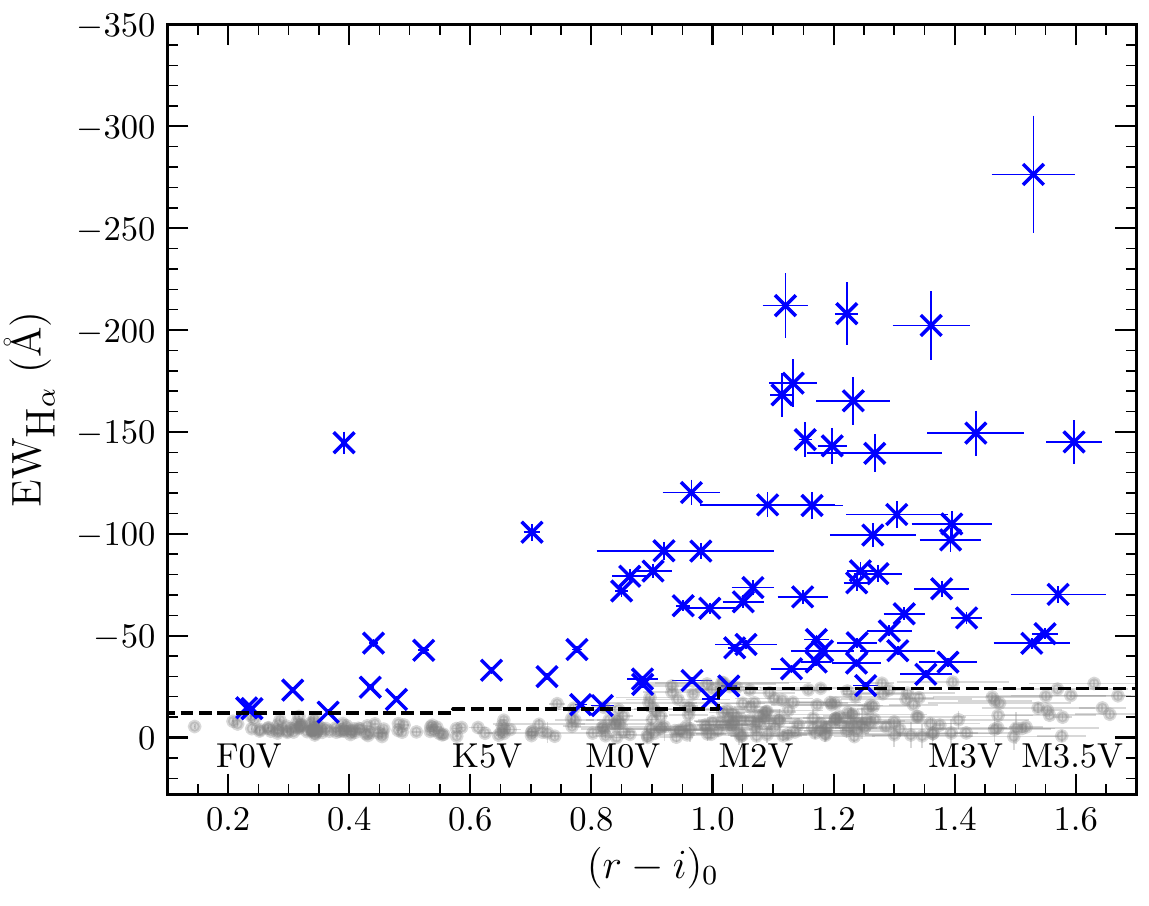}
\center
\caption{($r-i$)$_0$ vs. EW$_{\rm{H}\alpha}$ diagram of all sources. Stars selected as accretors are shown as blue crosses, along with their associated errors. The selection criteria is shown as a dashed line, and the spectral type at different values of ($r-i$)$_0$ are marked. }
\label{riew}
\end{figure} 

\subsection{Identification of accreting pre-main sequence stars based on photometric EW$_{\textrm{H}\alpha}$}

It has been established that the ($r-i$) vs. ($r-$H$\alpha$) diagram (Fig.\,\ref{riha}) can be used to obtain estimates of the H$\alpha$ equivalent width (EW$_{\rm{H}\alpha}$) of stars with precision photometry \citep{de10}, given the extinction and spectral type are relatively well-approximated (to within 0.2 mag). We follow the method detailed in \cite{kalari19} for VPHAS+ photometry to estimate the H$\alpha$ equivalent width of all VPHAS+ sources, adopting the average absolute visual extinction value ($A_V$) of 1.3\,mag from \cite{rho08}. 

From the values of photometrically estimated EW$_{\rm{H}\alpha}$, we select accreting PMS candidate stars based on their ($r-i$) colour (as a spectral type proxy). Note that ($r-i$) colours, and $r$-band magnitudes are corrected for H$\alpha$ emission, as the H$\alpha$ line falls within the $r$-band filter. Finally, images of accretors were also visually inspected using unsharp masking to remove sources affected by nebulosity. The reader is referred to \cite{Kalari15} for details on these procedures, and also the H$\alpha$ excess candidate selection criteria. To select PMS candidates, we use the spectral type-EW$_{\rm{H}\alpha}$ criteria for Classical T-Tauri stars (CTTS) of \cite{barr03}, adding to the upper limit of selected candidates based on the average error values of EW$_{\rm{H}\alpha}$ (CTTS are low-mass accreting PMS stars, showing H$\alpha$ in emission). Stars earlier than K5 spectral type having EW$_{\rm{H}\alpha}$ values less than $-12$\AA\ (where negative values denote in emission), in the K5-M2 spectral range with EW$_{\rm{H}\alpha}$ values less than $-$15\AA\,, and stars later than M2 with H$\alpha$ in emission less than $-$25\AA\ are selected as candidate PMS stars. Results of the selection procedure can be seen in Fig.\ref{riew}. The selection criteria account for propagated errors from the photometry, and uncertainty in spectral type. They are also large enough to weed out main-sequence contaminants with solely chromospheric H$\alpha$ emission. 68 PMS candidates were selected based on this method. The resulting EW$_{\rm{H}\alpha}$ values for the selected candidates, along with their photometry and astrometry is given in Table\,2.


\subsection{Kinematic properties of pre-main sequence candidates}

{\it Gaia} proper motions provide a valuable membership indicator for nearby star-forming regions. Given that the Trifid nebula is located in our line of sight towards the plane, it is likely that non-members may be erroneously identified as PMS members. Using proper motions, these can be excluded from further study.  

\begin{figure}
\center 
\includegraphics[width=8cm]{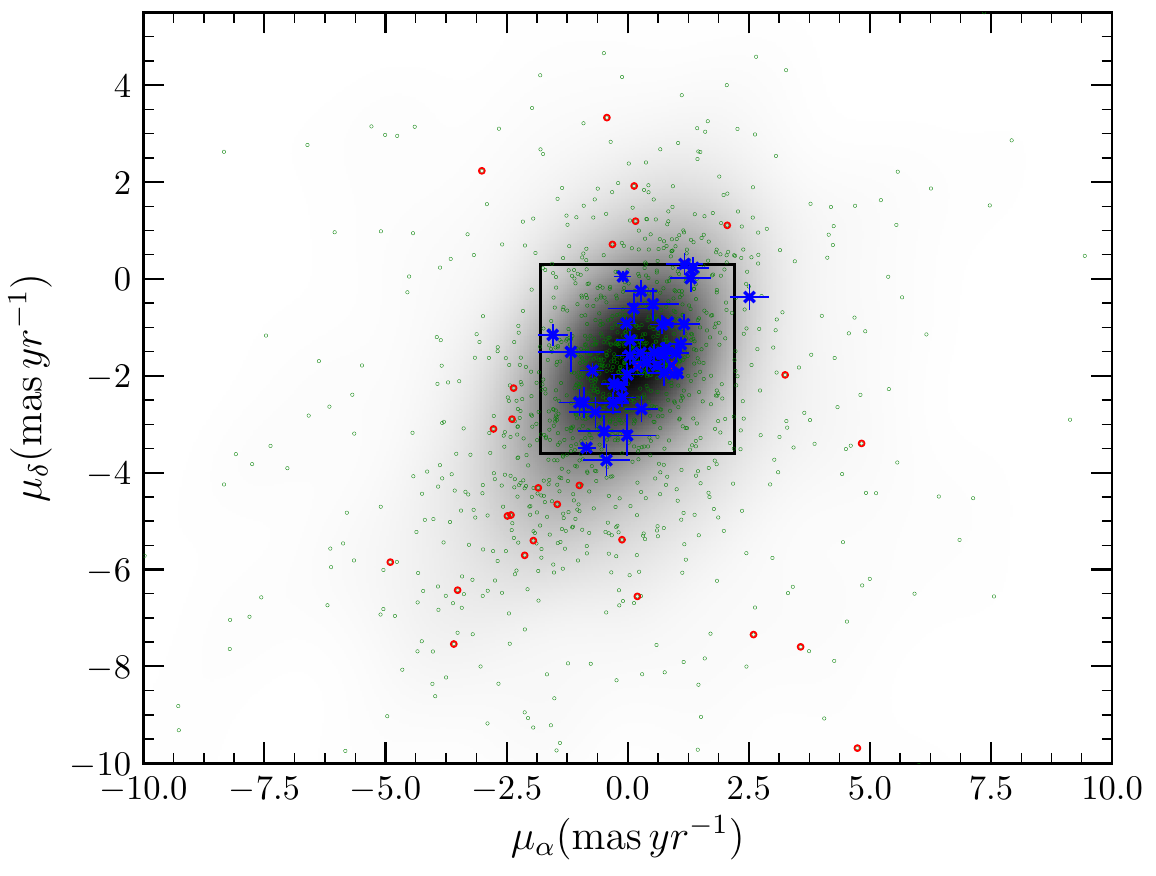}
\center
\caption{{\it Gaia} EDR3 proper motions of PMS candidates in the Trifid Nebula (green circles). The underlying inverted grayscale density map reflects the distribution of all {\it Gaia} sources in the field of view. Candidates selected as members on the basis of their proper motions are shown by blue crosses, while those rejected as members based on the proper motion criteria are shown as red circles. The proper motion selection criteria is represented by the solid bounding box. Dots represent proper motions of all stars in the area of study. }
\label{pm}
\end{figure}

M\,20 is identified in the $\mu_{\alpha}$--$\mu_{\delta}$ plane based on the mean motion of PMS candidates, having errors in any proper motion vector less than 0.5\,mas\,yr$^{-1}$. The cluster itself is separated in proper motion space with respect to the surrounding field population, with an increase in density visible around the cluster centre identified based on the motions of the PMS candidates (see Fig.\,\ref{pm}), and verified through the density map. Notice the large scatter of sources around the mean proper motion. This is expected given the location of the region in the Milky Way, as the photometry contains numerous background/foreground stars. The exact centre of the region can be best distinguished using the identified PMS candidates. 

To calculate the centroid in the proper motion plane, we fit double Gaussians to the proper motion vector histograms, which represent a conflation of the field population, having a broad distribution, and the cluster population, which has a well-defined peak. The proper motion centres in both directions was estimated based on the narrow Gaussian parameters. For $\mu_\alpha$, the centre is 0.22$\pm$0.8\,mas\,yr$^{-1}$, and in $\mu_\delta$ it is at $-$1.71$\pm$1.6\,mas\,yr$^{-1}$. The error bar denotes the width of the Gaussian peak. 

Proper motion candidates are selected based on the estimated proper motion centre of the cluster. We assume a median width of 2\,mas\,yr$^{-1}$ from the cluster centre. PMS candidates meeting the proper motion criteria are selected as those lying between $-1.8<\mu_{\alpha}<2.2$ and $-3.7<\mu_\delta<0.3$. Stars lying outside this bounding box are rejected as members, if their proper motions values along with the errors lie outside this boundary. 27 candidates were removed based on this criterion. Since the proper motion centre of the PMS members lies within the field proper motion distribution, it is possible that some non-members may have been included in our final analysis.

Finally, non-accreting members that can potentially be identified using {\it Gaia} proper motions are not included as the proper motion distribution of the cluster overlaps significantly with the field. A selection criterion to isolate such stars using only proper motions would include a non-negligible fraction of non-members, with no further defining physical characteristic to identify members.

\subsection{Distance to the Trifid nebula}

A large discrepancy exists in the literature on the distance towards M\,20. Previous distance measurements have been made using spectroscopic parallax \citep{rho08, tapia}, and extinction measurements \citep{cambresy11}. Spectroscopic parallax measurements of HD\,164492A place it at a distance between $\sim$1.6--2.0\,kpc \citep{rho08,tapia}, depending on the reddening law chosen (see \citealt{tapia}). This places it behind the Sagittarius arm, and in front of the Scutum arm of the Galaxy. Whereas distance measurements using extinction mapping by \cite{cambresy11} suggest a much farther distance of around 2.7\,kpc, placing it firmly in the Scutum arm of the Galaxy. In this paper, we eschew from depending on a priori knowledge of the reddening towards the region to measure the distances, by using high fidelity parallax values of identified PMS candidates. 

To estimate the distance towards the region, we consider only PMS candidates with {\it Gaia} EDR3 $\pi$ errors ($\sigma_\pi$) less than 0.1\,mas. As this criterion leaves us with only three stars, we add to the sample five previously identified members from the YSO sample, which all have $\sigma_\pi<$\,0.1\,mas. This increase from the IR identified YSO sample is expected as that sample is concentrated towards brighter magnitudes, where the $\sigma_\pi$ are lower. In total, we have 8 stars that are candidate members of the region, with high-fidelity $\pi$. Our chosen error criteria is to remove faint stars with non-negligible errors, that may degrade the accuracy and precision of the final distance estimate. 

\begin{figure}
\center 
\includegraphics[width=8cm]{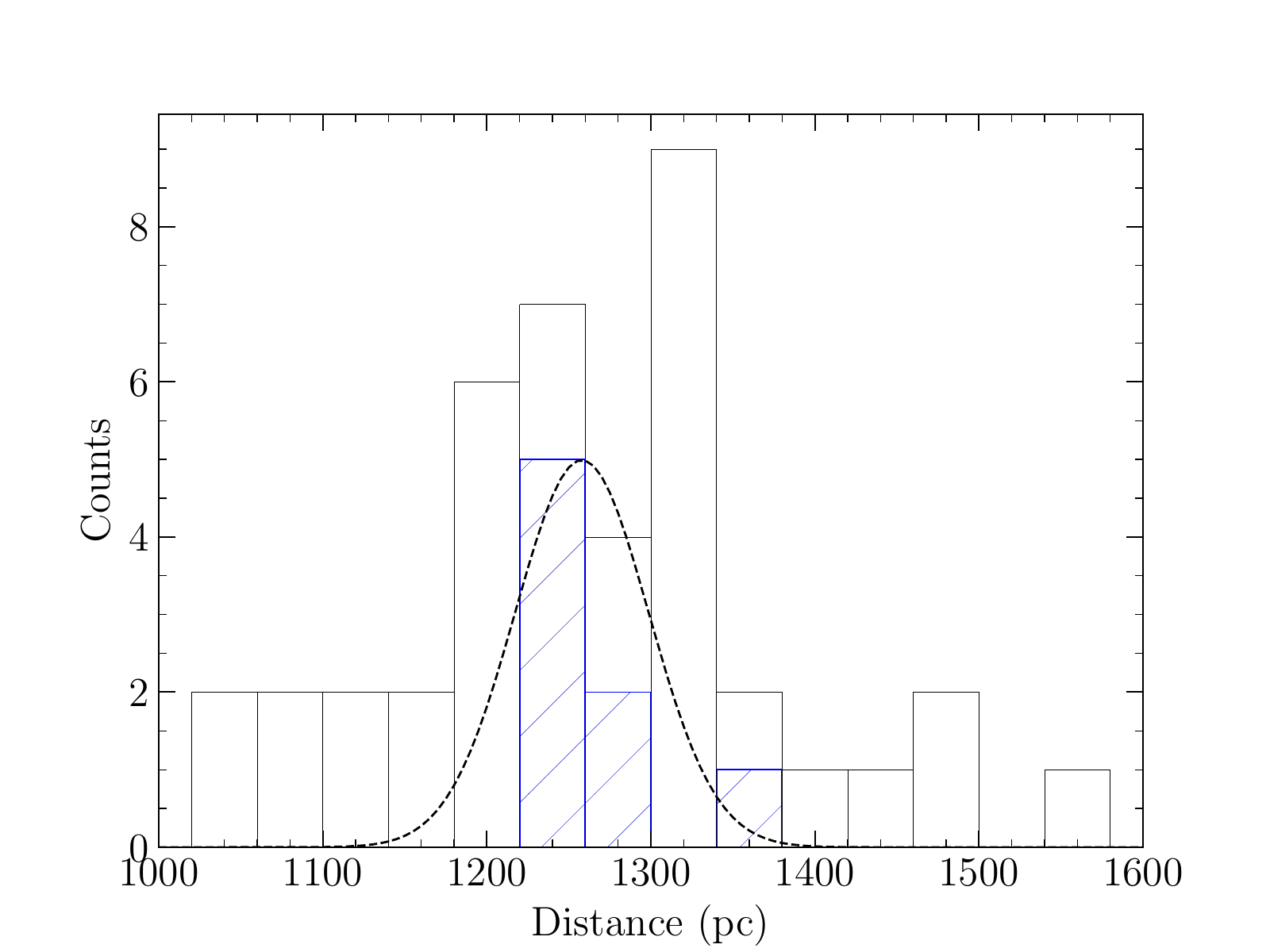}
\center
\caption{Histogram of distances of candidate members having $\sigma_\pi<$0.1\,mas (hatched bars). The resulting Gaussian is shown as a dashed line, and has a peak of 1250\,pc. Also shown only for comparison are the distances of all {\it Gaia} EDR3 sources meeting the proper motion membership and parallax error criteria. }
\label{dist}
\end{figure}

To estimate the individual distances of the chosen sample, we used the method of \cite{bailerjones}. An inference approach is adopted, which uses a prior with a length scale parameter (chosen here to be 1750\,pc, based on the Galactic model given in \citealt{bailerjones}). The resulting confidence intervals of the individual distances are asymmetric. Based on this method the resulting distance histogram of all candidates is shown in Fig.\,\ref{dist}. An inverse weight of the mean error was applied to the histogram. A Gaussian is fit to the histogram, accounting for the well defined cluster distance. From the peak of the Gaussian fit, we estimate a distance of 1257$^{+190}_{-98}$\,pc to M\,20 (the median distance of the sample is 1255\,pc). The distance towards the cluster is taken to be 1250\,pc for our analysis. The asymmetric error represents the median of the 5th and 95th percentile confidence intervals of the resultant distance distribution. 

The resulting distance from {\it Gaia} EDR3 parallaxes is much closer than previous measurements using spectroscopic parallaxes or extinction mapping. It agrees with the estimate of $\sim$1.22\,kpc made using {\it Gaia} DR2 parallaxes by \cite{wade}. As a comparison, also shown are the histogram of distances of {\it Gaia} EDR3 sources meeting the proper motion criterion, and having RUWE and $\sigma_\pi$ within the limit for our sample. This sample has a median distance of 1300\,pc, but may contain few contaminant background stars. Finally, the new closer distance places the cluster firmly in the Sagittarius arm, alongside the star forming regions in the near vicinity such as the Lagoon Nebula, Simeis\,188 (both lying below the Galactic plane), and the optically invisible H\,{\scriptsize II} region W28 A2, and its supernova remnant W28. 

\subsection{Stellar properties of pre-main sequence candidates}

\begin{figure}
\center 
\includegraphics[width=8cm]{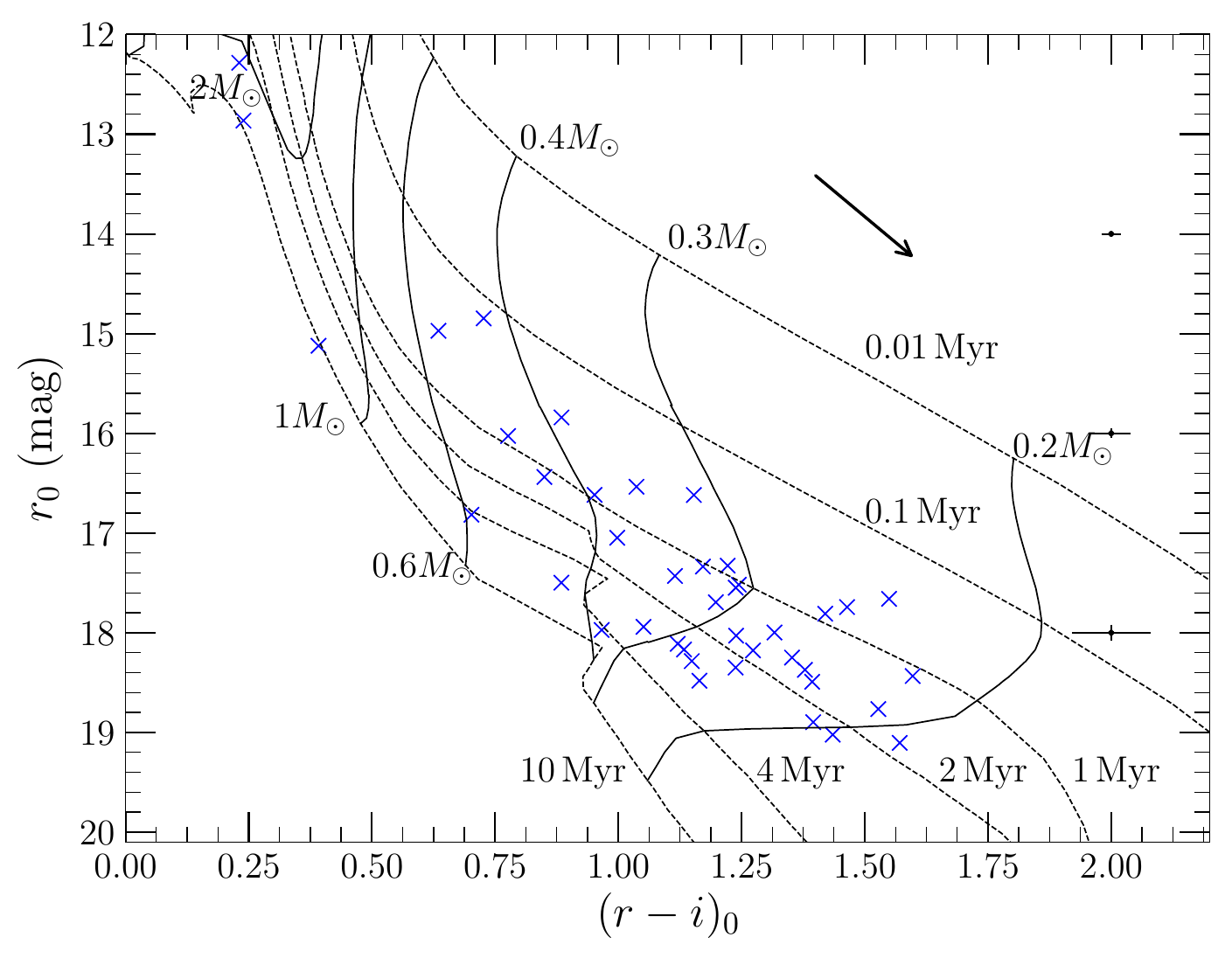}
\center
\caption{($r-i$) vs. $r$ colour-magnitude diagram of PMS candidates. Overplotted and labelled are the isochrones, and mass tracks from \citep{bress12}. The extinction vector for $A_V=1$\,mag is shown in the top right.  }
\label{cmd}
\end{figure}

We estimate the stellar properties (mass, $M_\ast$ and age, $t_\ast$) of the 41 PMS candidates that meet the proper motion criteria, by interpolating their positions in the observed ($r-i$) vs. $r$ colour-magnitude diagram compared to model isochrones and mass tracks (Fig.\,\ref{cmd}). The stellar isochrones and mass tracks of \cite{bress12} in the VPHAS+ filter set (AB magnitude system) are utilised for this purpose. The AB to Vega offset values for the $r$ and $i$-band filters were 0.174 and 0.378 respectively. To understand the differences between the choice of stellar models, the $M_\ast$ and $t_\ast$ are compared to those estimated using the \cite{siess00} and \cite{dotter} models in the VPHAS+ filter set. The metallicity in the models is set to solar, and the models only account for single stars. The $A_V$ is assumed to be uniform, and is set at 1.3\,mag, with an $R_V$ of 5.5 \citep{cambresy11} used to deredden the observed photometry. A distance of 1250\,pc was used to transform the models to the observed distance. The results are reported in Table\,2. Errors on the interpolated stellar properties include propagated photometric uncertainties, and an assumed distance and extinction uncertainty of 200\,pc, and 0.2\,mag respectively. The effect of assuming a uniform extinction value on our results is also discussed.

\begin{deluxetable*}{lllllcclc}
\tablenum{2}
\tablecaption{Optical photometry, {\it Gaia} astrometry, and estimated stellar properties of pre-main sequence stars in the Trifid Nebula. \label{tab:messier}}
\tablewidth{0pt}
\tablehead{
\colhead{VPHAS+ ID} & \colhead{$r$} & \colhead{$i$} & \colhead{H$\alpha$} & \colhead{EW$_{{\rm{H}}\alpha}$} &
\colhead{$\mu_{\alpha}$} & \colhead{$\mu_{\delta}$} & 
\colhead{Mass} &\colhead{Age} \\
\colhead{} & \colhead{(mag)} & \colhead{(mag)} & \colhead{(mag)} &\colhead{(\AA)} & \colhead{(mas\,yr$^{-1}$)} & \colhead{(mas\,yr$^{-1}$)} & \colhead{($M_{\odot}$)} & \colhead{(Myr)}
}
\startdata
  816177248012 & 20.07$\pm$0.07 & 18.53$\pm$0.03 & 18.53$\pm$0.06 & $-$149.42$\pm$11.1 & $-$1.173$\pm$0.678 & $-$1.508$\pm$0.412 & 0.19$^{+0.01}_{-0.02}$ & 2.24$^{+0.39}_{-0.44}$\\
    816177242912 & 13.41$\pm$0.01 & 12.99$\pm$0.01 & 13.11$\pm$0.01 & $-$14.56$\pm$1.5 & $-$0.011$\pm$0.596 & $-$3.23$\pm$0.433 & 1.88$^{+0.36}_{-0.23}$ & 6.08$^{+2.73}_{-3.00}$\\
      816177256886 & 20.19$\pm$0.07 & 18.47$\pm$0.03 & 18.96$\pm$0.07 & $-$70.25$\pm$4.1 & 0.522$\pm$0.544 & $-$0.522$\pm$0.342 & 0.18$^{+0.01}_{-0.02}$ & 1.81$^{+0.28}_{-0.35}$\\
  816856488067 & 19.55$\pm$0.044 & 18.25$\pm$0.025 & 18.23$\pm$0.035 & $-$113.88$\pm$6.5 & $-$0.667$\pm$0.538 & $-$2.75$\pm$0.319 & 0.25$^{+0.01}_{-0.02}$ & 3.01$^{+0.68}_{-0.8}$\\
  816856492037 & 19.87$\pm$0.05 & 18.18$\pm$0.02 & 18.79$\pm$0.05 & $-$46.22$\pm$2.6 & 0.123$\pm$0.529 & $-$0.609$\pm$0.313 & 0.21$^{+0.01}_{-0.01}$ & 1.58$^{+0.26}_{-0.37}$\\
  \enddata
\tablecomments{Five selected rows are shown to display the form and contents of the table. The full table is only available electronically, along with the source coordinates. Mass and age errors denote the upper and lower deviations based on the propagated photometric uncertainties, and assuming a distance and reddening uncertainty of 200\,pc, and 0.2\,mag. }
\end{deluxetable*}

\begin{figure}
\center 
\includegraphics[width=8cm]{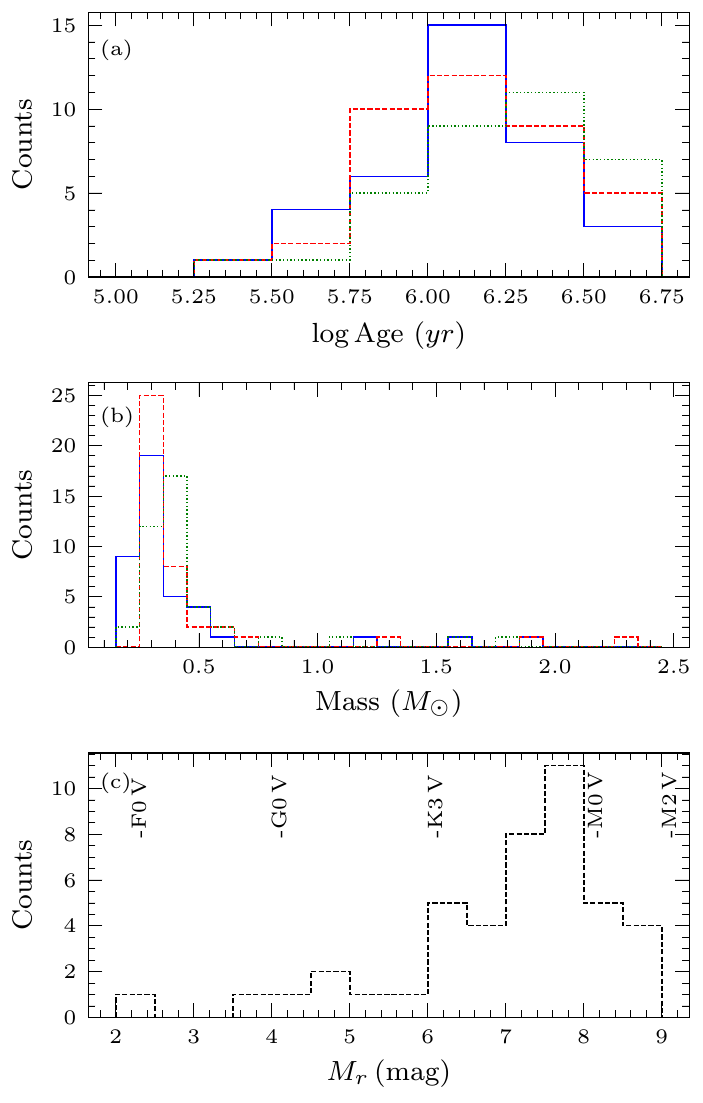}
\center
\caption{Age, Mass, and $M_r$-band histograms of PMS candidates in the Trifid Nebula in panels ({\it a}), ({\it b}), and ({\it c}) respectively. For the age, and mass histograms, the solid (blue), dotted (green), and dashed (red) lines represent values estimated using the \cite{bress12}, \cite{dotter}, and \cite{siess00} models respectively. In the $M_r$ histogram the approximate absolute $r$-band magnitude of select spectral types from \cite{mamajek} are labelled. }
\label{hist}
\end{figure}

The resulting colour-magnitude diagram of the PMS stars, with the \cite{bress12} stellar tracks and isochrones overlaid is shown in Fig.\,\ref{cmd}. PMS candidates from our sample fall approximately between the 0.2--2.0\,$M_{\odot}$ tracks, and are clustered between the 1 and 2\,Myr isochrones. The results estimated from the three different stellar models are given in Fig.\,\ref{hist}. Here, the median mass of our candidates peaks around 0.3\,$M_{\odot}$, with the most massive stars $\sim$ 2\,$M_{\odot}$. This mass distribution is expected given the absolute $r$-band luminosity function of our candidates (Fig.\,\ref{hist}c), where our candidates fall between 2--9\,mag (after correcting for extinction and distance). This approximates to early F-- early M candidates, which corresponds to the derived mass range. Also marked for reference in Fig.\,\ref{hist}c are the expected absolute $r$-band magnitudes for main-sequence stars. There are no significant differences between stellar masses due to the assumption of a particular set of models accounting for standard errors. Note that the mass of the most massive stars according to the \cite{siess00} models is slightly higher when compared to the other two model results.

The median age of the PMS candidates interpolated using the \cite{bress12} isochrones is 1.5$^{+0.3}_{-0.5}$\,Myr. Error bars correspond to the median values for the upper and lower age errors, which are estimated from interpolated ages including the photometric, reddening, and distance uncertainties. The age distribution of the PMS candidates is shown in Fig.\,\ref{hist}a. Note that the reddening vector exhibits a similar slope to the isochrones in the ($r-i$) vs. $r$ plane (Fig.\,\ref{cmd}). Minor increments in extinction will not change significantly the estimated age of the region, as evidenced by the resulting age errors.
The median ages from the \cite{bress12}, \cite{siess00}, and \cite{dotter} models are 1.5, 1.4, and 1.9\,Myr respectively, and that the estimated age differences between models are lower than the standard errors. For the remainder of this paper, we utilise the stellar properties derived using the \cite{bress12} models. 

The median age of the PMS stars estimated ($\sim$1.5\,Myr) here is older than the literature age of the region $\sim$0.3\,Myr. This age estimate was made based on the expected wind-bubble shell size by \cite{Cernicharo}. Although this value does not take into account earlier stages, it represents a good lower limit to the cluster age. The photo-ionisation time of cometary globules in the region \citep{lefloch}, and the presence of dense star-forming cores \citep{tapia} attest to a young age ($\lesssim$1\,Myr), as being the most recent burst of star formation in the region \citep{rho08}. The age estimated in this work represents well only the current generation of PMS stars in the region. This generation of stars must have formed alongside or slightly before HD\,164492A, as its spectroscopic age is around 0.6\,Myr \citep{164492Age}.

The impact of our assumption of a uniform extinction value $A_V$=1.3\,mag on our resultant ages is tested by adopting higher uniform values of extinction. Assuming $A_V$=2\,mag, increases the median age of the sample slightly to $\sim$1.8\,Myr. This agrees with the expectation of the reddening vector discussed earlier. Assuming a higher extinction of $A_V$=5\,mag, around 10 stars fall beyond the 100\,Myr isochrone, and their ages cannot be estimated. These stars would be considerably blue, and lie beyond the main-sequence at the distance of 1250\,pc. The median age of the remaining sample is 2.8\,Myr. Thus, it is likely that the PMS stars have a lower mean extinction than assumed here, than higher. Finally, we found that a simulation of the random variation of individual stellar reddening for PMS stars leads to an age estimate of 1.7\,Myr. To arrive at this value, we generated random values for the assumed extinction correction (between $A_V$=0.5 and 5\,mag) for each star. Each star was also cloned 10000 times, and the median age of these clones with random extinction correction was the result. These experiments serve to demonstrate that the assumed mean extinction is not a significant under-estimation for the majority of our sample; and that while individual stellar extinction values are essential for a precise age estimate, the final median age of our sample should not vary significantly. 
From this analysis, we also suggest that future spectroscopic surveys of the young stellar population that can estimate precise individual extinctions and spectroscopic properties can pin down the precise age of individual stars in the Trifid nebula.

\begin{figure*}
\center 
\includegraphics[width=16cm]{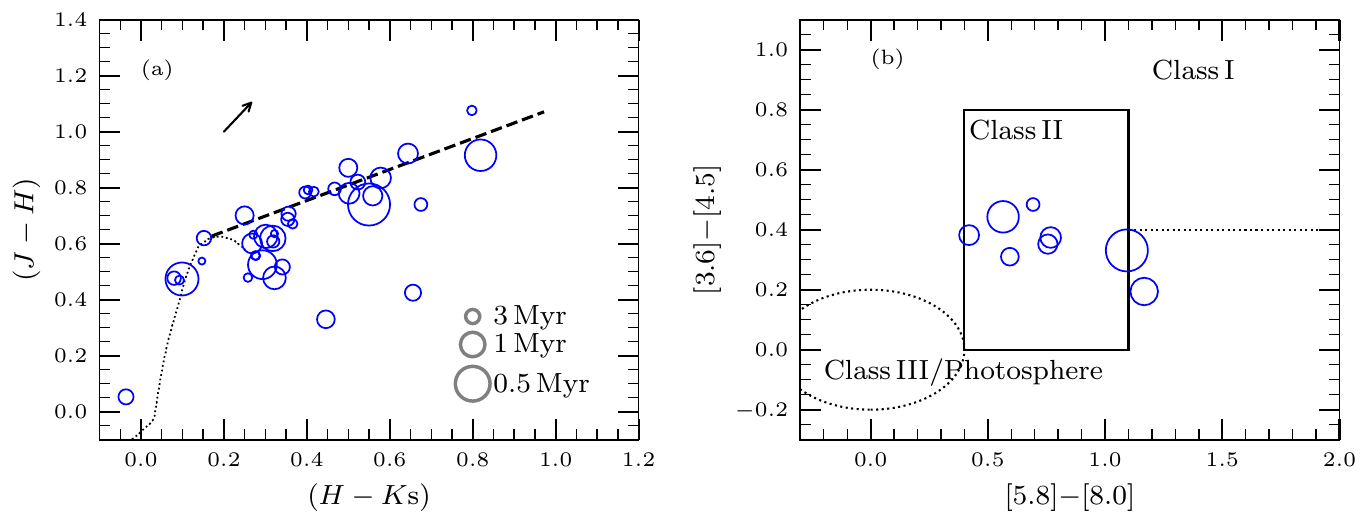}
\center
\caption{({\it a}): ($H-K$s) vs. ($J-H$) colour-colour diagram of the PMS sample. The reddening vector for $A_V$\,=1\,mag is also shown in the top left. The CTTS locus from \cite{meyer97} is shown as a dashed line. The main-sequence colours from \cite{mamajek} are shown by a dotted line. ({\it b}): The mIR colours of the sample are shown. The boundary of colours for Class II sources with a median accretion of 10$^{-8}$\,$M_{\odot}\,yr^{-1}$ from the models of \cite{d05} are shown, along with the positions of Class I and Class III sources. In both panels, the the size of the symbol is scaled with the estimated stellar age, given in the index in the lower right of panel (a). }
\label{irac}
\end{figure*} 

\subsection{Infrared properties of pre-main sequence candidates}

As described in Section\,3, circumstellar discs surrounding PMS stars are heated by the central star. This leads to an inner disc wall which re-radiates and absorbs energy primarily at nIR wavelengths leading to prominent $K$-band excesses compared to main-sequence stars. Evidence for nIR--mIR excesses in PMS stars are thus excellent independent indicators of circumstellar discs, and accretion. The exact value of the IR excesses depends on many factors \citep{meyer97}. Note that an absence of excess does not conclusively rule out the presence of a cirumstellar disc, but could be due to an inner disc hole, or edge-on disc inclination angle. 

To identify if our candidate PMS stars have near or mIR excesses resembling circumstellar discs, we use the cross-matched nIR and mIR data described in Section\,2.4. 39 PMS candidates have nIR counterparts in $JHK$s bands from the UKIDSS survey meeting the photometric criteria. The ($H-K$s) vs. ($J-H$) colour-colour diagram is shown in Fig\,\ref{irac}a. The main-sequence locus from \cite{mamajek} is shown along with the extinction vector for 1\,mag. Also marked is the CTTS locus from \cite{meyer97}, showing the expected nIR excesses for CTTS stars having accretion rates between 10$^6$--10$^8$\,$M_{\odot}\,yr^{-1}$. 90\,\% of our sample have colours within errors falling on the CTTS locus, suggestive that the majority of our sample have circumstellar discs. The lack of nIR excesses for a few stars in our sample does not suggest the absence of circumstellar discs, and could be due to inclination angle or inner disc hole.

While 32 PMS candidates have counterparts in one or more mIR {\it Spitzer} band, only eight have photometry in all bands and are shown in Fig.\,\ref{irac}b. Seven are marked as Class II YSOs from the $\alpha$ analysis (with the remaining star having $\alpha=-1.9$). For the GLIMPSE survey, the limiting magnitude in the {\it Spitzer} [3.6] band is around 14\,mag, which is approximately a mid-K spectral type. Therefore, it is not surprising that while nearly all our candidates have nIR counterparts with high quality photometry from UKIDSS, only a brighter sub-sample have mIR photometry. The expected colours for stars having mean accretion rates of 10$^{-8}$\,$M_{\odot}\,yr^{-1}$ from \cite{d05} is shown, along with the expected positions of Class\,I and Class\,III sources. All of the sample falls in the expected positions for Class II sources with the mean accretion rate around 10$^{-8}$\,$M_{\odot}\,yr^{-1}$. The nIR and mIR colours of our sample clearly suggest that they are PMS stars. The IR colour-colour diagrams shown in Fig.\,\ref{irac} also relates the position in colour space to the estimated age. No clearly demarcated trend is seen from the observed colours. Stars that are detected at mIR wavelengths are younger than the general sample. The photometry of the PMS sample in nIR and mIR is given in Table\,2.

\subsection{Sample completeness}

\begin{figure}
\center 
\includegraphics[width=8cm]{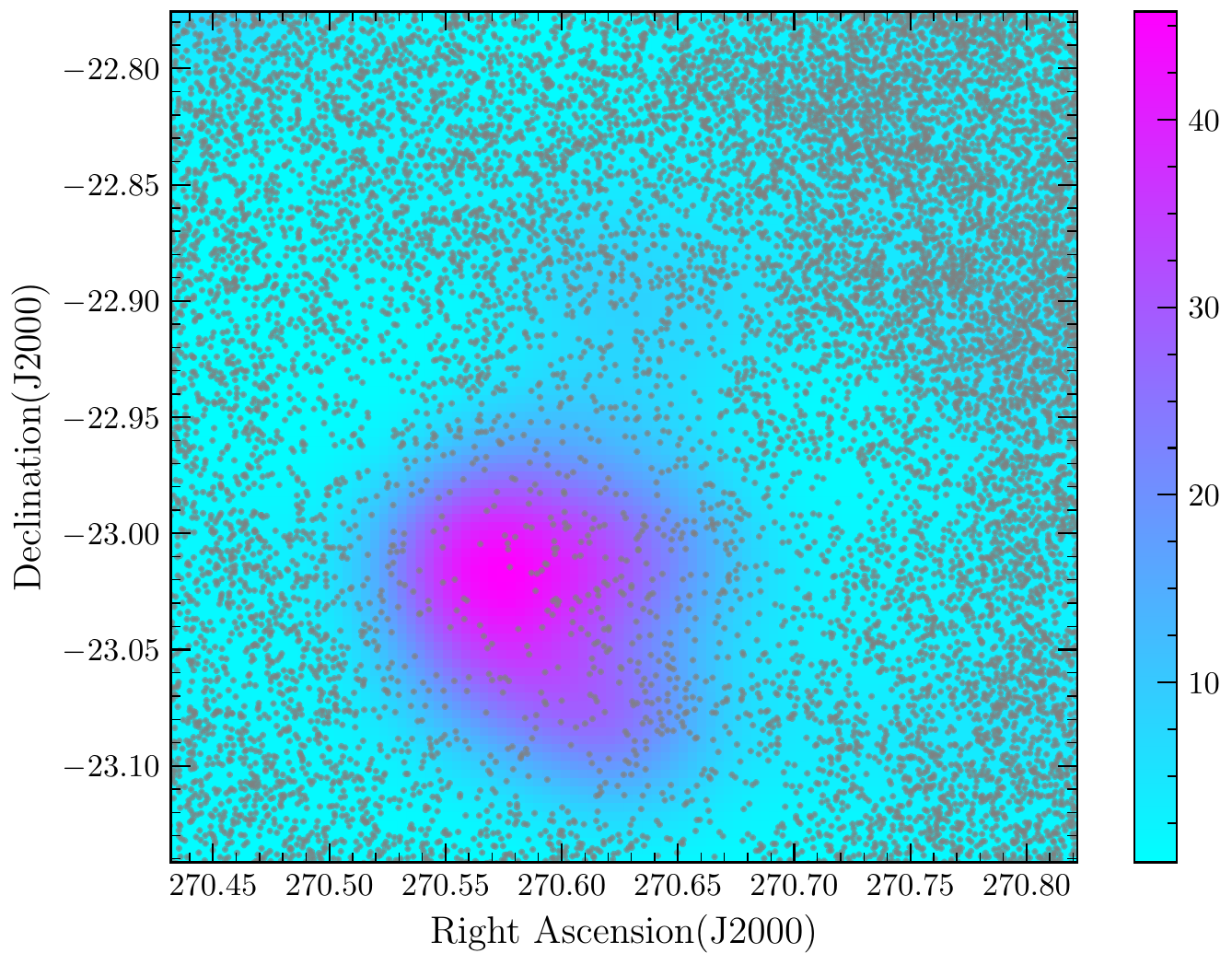}
\center
\caption{Colour map in right ascension and declination depicting ratio of sources detected to those whose photometry is affected by spatially varying nebulosity on small scales. {\it Gaia} EDR3 sources are overlaid as grey circles. }
\label{spacecompare}
\end{figure}

Sample completeness is a function of photometry meeting our quality criterion for all sources in the Trifid nebula in magnitude and spatial location. Completeness in magnitude space estimated from the luminosity function in $ri$H$\alpha$ are 19.8, 18.5, and 19.4\,mag respectively (see \citealt{Kalari15} for details). Based on the photometric magnitudes, we are complete until a stellar mass around 0.45\,$M_{\odot}$. Spatial variation of completeness is considerable in our sample due to nebulosity varying on small spatial scales. In this scenario, high-fidelity photometry of sources cannot be determined using currently available sky subtraction techniques (see Section 2.2). Note that while these sources are identified, their photometry particularly in H$\alpha$ is sufficiently degraded to be unsuitable for further analysis. To gauge this incompleteness, we estimate the fraction of the total number of sources detected in the $ri$H$\alpha$ magnitude across the spatial extent of the study, to the number of sources meeting the photometric quality criteria in equally spaced grids. This fraction is represented in Fig.\,\ref{spacecompare}. This is created by assuming that for an increasing aperture radius, if the increase in observed magnitude is larger than the error on the magnitudes, the photometry is affected by the increasing nebulosity (i.e., being classed as an extended source in \citealt{drew14}). This is given by the equation
\begin{equation}
    |r_{\textrm 4}-r_{\textrm 3}|>3\times\sqrt{{r_{\textrm 4}}^2+{r_{\textrm 3}}^2}+0.05.
\end{equation}
Here $r_{\textrm 4}$ and $r_{\textrm 3}$ represent the magnitudes in apertures of 2.8$\arcsec$ and 2$\arcsec$ respectively. The smoothed density map of sources affected by nebular subtraction in Fig.\,\ref{spacecompare} suggests that the sample is more incomplete towards the central dust lane, and HD\,164492A. Accounting for the missing fraction of PMS stars due to this spatial incompleteness is difficult, as it depends on the inherent spatial distribution, and their fraction compared to the total number of sources in a given region. This caveat is considered when analysing the spatial distribution in the Trifid in the region around HD\,164492A. Also over plotted in Fig.\,\ref{spacecompare} is the spatial distribution of {\it Gaia} EDR3 point sources, as a comparative reference for high-quality optical imaging in the region.

Overall, we can conclude that our sample is complete down to masses of 0.45\,$M_{\odot}$, and spatially our sample is significantly affected by the nebulosity. The effect of this incompleteness suggests sources missing around the central dust lane, and near HD\,164492A. Based on this, we can assume that our sample represents only a lower limit of PMS stars around HD\,164492A. Given these circumstances, we define a radius around the central O star where the sample is incomplete, and consider this caveat when discussing the spatial properties of the PMS sample.

\section{Discussion}
 
\subsection{Star formation triggered by CCC}

\begin{figure*}
\center 
\plotone{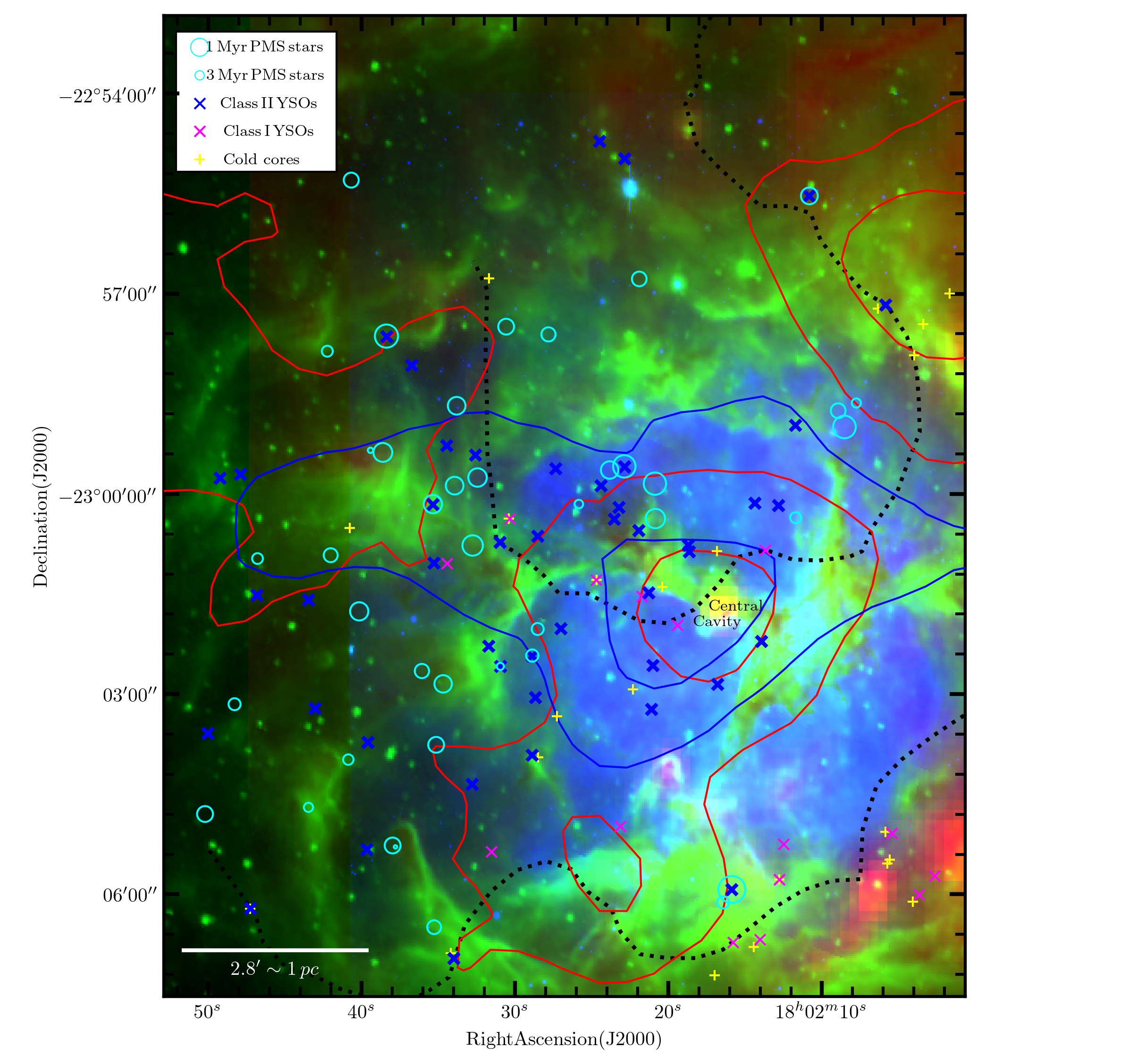}
\center
\caption{{\it Herschel} 350$\mu$m/8$\mu$m {\it Spitzer}/VPHAS+ H$\alpha$ $rgb$ image of the Trifid Nebula. North is up and east is left. The scalebar corresponding to 1\,pc at the distance of the Trifid nebula is also given. Circles mark the location of PMS stars detected in our sample. Their size is related to their age, and scaled according to the legend in the top left corner. Blue and cyan crosses mark the position of Class\,II and Class\,I sources respectively, while the plus sign indicates the location of star-forming cores from \cite{tapia}. The blue and red contours indicate the blue-shifted 2\,km\,s$^{-1}$ cloud, and red-shifted cloud C which have collided following \cite{torii11}. The contour levels are drawn for clarity at 18 and 36\,K\,km\,s$^{-1}$ for both clouds. Black contours represent the filaments identified from {\it Herschel} imaging. }
\label{moneyplot}
\end{figure*}

As discussed in Section\,1, the Trifid Nebula is a proposed site of CCC triggering star formation. From a sub-mm study of CO lines \cite{torii11, torii17} identified three distinct velocity components at 1.4, 9 (here after 2\,km\,s$^{-1}$ cloud, and cloud C respectively following their nomenclature), and 18.2\,km\,s$^{-1}$, representing three different molecular clouds. They claimed that the 2\,km\,s$^{-1}$ cloud, and cloud C are observed to have higher temperatures, and are likely heated by the central ionising source (HD\,164492A) based on positional and luminosity arguments-- suggesting that they are the parent clouds of the central cluster (while the 18\,km\,s$^{-1}$ cloud is unlikely to be related to current star formation in the region). \cite{torii11} indicate that the observed velocity pattern is systemic. The 2\,km\,s$^{-1}$ cloud is presently moving towards us, with cloud C moving away from us with an observed velocity separation of 7.5\,km\,s$^{-1}$. Here, the velocity separation refers to the differences in the peak velocity of the two clouds CO spectra, and is taken from \cite{torii11}. The 2\,km\,s$^{-1}$ cloud is associated with the dark lanes in the Trifid nebula, but cloud C does not correspond to the optically observed features strengthening the argument that cloud C lies behind the 2km\,s$^{-1}$ cloud in our line of sight. The total estimated molecular and stellar mass is insufficient to gravitationally bind the system, leading the authors to propose a different mode of star formation. They suggest that a consistent scenario with the observed velocity and spatial properties of the molecular clouds is that these two clouds collided $\sim$1\,Myr ago, with the observed relative velocity difference arising from the collision (see also Fig.\ref{moneyplot}). The collision between these two clouds triggered rapid star formation at the location of the collision, including the formation of HD\,164492A.

Important differences between stars forming in a single molecular cloud versus a pair of colliding clouds are the spatial positions and age spreads of the newly formed stars with respect to the position and velocities of their natal molecular clouds \citep{loren1976}. When gravitational collapse of molecular clouds is induced by turbulence, it is thought that heavier filaments fall towards the centre, and the majority of stars in the central region form over a prolonged period of time from such dense material giving rise to a concentration of young stars in cluster centres, with the more massive stars forming there preferentially. Older stars are expected to be detected more towards the outskirts of regions either due to lower molecular densities in the outer areas, resulting in the earliest period of star formation exhausting material to form stars, or due to dynamical interactions over time. In the case of head-on CCC as considered in M\,20, the new generation of stars will form where the densities and compression are the highest. This is expected to be towards the edges of the collided clouds, as a cavity is created towards the centre of the collision, along with a detected bridge feature at intermediate velocities \citep{torii17}. The stars which formed rapidly after the collision will now be found in a region where the two separate velocity components are coalescing into one intermediate velocity. The younger generation of star formation currently occurring will be concentrated towards the edges of the collided clouds, and near the edges of the cavity where dense molecular material is not yet exhausted. Stars formed as a result of the collision will be found exclusively around the collapsing region, but the central overlapping region where the clouds collided might be devoid of stars.

Therefore, the spatial and age spreads of young stars can be correlated with the velocity and spatial properties of molecular clouds and filaments to identify the sites of CCC induced star formation. Based on the demographics of the young stellar population identified in Sections 3 and 4, we explore here whether they provide evidence for or against the CCC hypothesis.

\subsubsection{State of colliding clouds in Trifid nebula}

The mass of the two clouds is calculated based on the CO luminosities and line widths reported in \cite{torii11}, where the authors adopted the $X_{\rm CO}$ factor of 2.0\,$\times$\,10$^{20}$\,cm$^{-2}$(K\,km\,s$^{-1}$)$^{-1}$ to convert the CO luminosity to mass. The difference between the calculations reported here and those in \cite{torii11} is in the distance adopted. For the calculations reported in this paper the distance adopted is 1250\,pc from {\it Gaia} EDR3 parallaxes, wheras \cite{torii11} adopted much larger distance estimates of 1700 and 2700\,pc. Based on the new distance, the total mass and radius of cloud C is estimated as 0.5$\times10^3$\,$M_{\odot}$ and 0.9\,pc respectively. The 2 km\,s$^{-1}$ cloud has a mass of 0.35$\times10^3$\,$M_{\odot}$ and a radius of 0.7\,pc. Considering the gravitational situation of the cloud, we can calculate whether it is bound or unbound based on the $\alpha_{\rm G}$ parameter. If $\alpha_{\rm G}>1$ the cloud is considered unbound, and if $\alpha_{\rm G}<1$ the cloud is bound. Here,
\begin{equation}
    \alpha_{\rm G} = \frac{5\sigma_v\,R}{GM}.
\end{equation}
The projected distance and velocity separation (the actual observed separation maybe higher depending on the inclination angle) between the clouds is denoted by $R$ and $\sigma_v$ respectively. From \cite{torii11}, we adopt $\sigma_v$ of 7.5\,km\,s$^{-1}$ and $R$ of 1--2\,pc. From this, we estimate that the mass, $M$ required to bind the clouds is $\sim$10$^4\,M_{\odot}$, an order of magnitude higher than both the stellar and cloud masses combined \citep{torii11}. Hence the clouds can be considered to be gravitationally unbound. Based on their velocity difference, they are expanding away from each other. Note that \cite{torii11} arrived at a very similar conclusion, albeit for a slightly higher distance estimate. Similarly, \cite{torii11} calculated that the mechanical luminosity generated by the stellar wind of HD\,164492A is insufficient to cause to the observed cloud expansion. 
Therefore, we can conclude similar to \cite{torii11} that the two molecular clouds are expanding away from each other and are not gravitationally bound, and the observed cloud velocities must be systemic. Based on the present direction of the velocities and the cloud size, the two clouds must have collided $\sim$0.7\,Myr ago \citep{torii11}, not accounting for projection effects. This value may vary based on projection effects by a factor of 2.

The dynamical age ($t_{\textrm{dyn}}$) of the H\,{\scriptsize II} region encompassing the Trifid nebula is estimated following;
\begin{equation}
    t_{\rm{dyn}} = \frac{4R_s}{7c_s}\left[ \left(\frac{R_{{\rm H{\scriptsize II}}}}{R_s}\right)^{(7/4)}-1 \right].
\end{equation}
$R_s$ is the radius of the Str{\"o}mgren sphere, $R_{\rm H{\scriptsize II}}$ is the radius of the Trifid nebula, and $c_s$ is the sound speed. $R_s$ is also calculated following eq.\,(1) in \cite{tremblin}. To calculate $R_s$, a recombination coefficient of 3$\times10^{-13}$\,cm$^3$\,s$^{-1}$ (adopting a temperature of 8000\,K), and a density of 3400\,cm$^{-3}$ is assumed from \cite{tremblin}, as it best represents the solar neighbourhood. We also adopt the ionising flux from \cite{smith02} for the central O7.5V star to be 5$\times10^{48}$\,photons\,s$^{-1}$. Differences in the assumptions of mean density can critically change the dynamical age (e.g. \citealt{tremblin}), so we calculate errors bars on our sample assuming a mean density range of 1000--10000\,cm$^{-3}$. From Fig.\,1, we estimate $R_{\rm H{\scriptsize II}}\sim$2.5\,pc, assuming a distance of 1250\,pc. The resulting dynamical age is $\sim$0.85$^{+0.15}_{-0.45}$\,Myr. This is larger than the estimate of 0.3\,Myr arrived at by \cite{Cernicharo}, assuming a distance of 1680\,pc, and density of 1000\,cm$^{-3}$. The value reported is in agreement with the spectroscopic age of HD\,164492A ($\sim$0.6\,Myr).

\subsubsection{Filamentary structure}

Long-wavelength images of the Trifid Nebula taken from {\it Herschel} space telescope display long filamentary structures. These structures can be identified and extracted using various decomposition algorithms. To identify the filamentary structures visible in the 500$\mu$m {\it Herschel} image of the Trifid Nebula, we utilise the FilFinder algorithm from  
\cite{koch15}. The `filaments' are identified from a flattened and masked image to remove bright compact sources, and pick out the filamentary structure. The size threshold for masking chosen was experimented with to better highlight the small scale and faint regions when visually comparing to the flattened image, and a value of 600 was used. The final pruned skeletons representing the `filaments' are shown in Fig.\,\ref{moneyplot}. From the image, we identified two filaments in the Trifid nebula. The two structures do not correspond visually to the optical nebulosity, and the three visible dust lanes. The filaments run along the visible emission in the mIR emission, but do not intersect. 

A clear distinction can be made between the two filaments in terms of the young stellar population. Along the northern filament lies the bulk of the Class\,II YSO and PMS population detected, while only three fall along the southern filament. The southern filament, especially towards the south west edge falls across a clump of cold cores identified by \cite{lefloch}, and a grouping of Class\,I YSOs. This suggests that stars forming along the southern filament must be younger than the bulk of the stellar population identified lying along the northern filament given the relative ratio of PMS stars and Class\,II YSOs to Class\,I YSOs and cold cores.

\subsubsection{Spatial distribution and star formation chronology of YSOs and accreting PMS stars}

The mean ages of the Class\,II YSOs are assumed to be similar or slightly younger than the accreting PMS stars, whose median age is 1.5\,Myr. The Class\,I YSOs are considered to be much younger \citep{harvey}, around $\sim$0.5\,Myr. The dynamical age of the region ($\sim$0.85\,Myr) is larger than the estimated age of the Class\,II YSOs, with the former specifying a lower limit to the age of the region. Further precise estimates of the reddening and distance may verify the ages of the accreting PMS stars and Class\,II YSOs (see Section 4.4). An additional timescale in consideration here is the crossing time. Following \cite{reviewadamo}, and assuming a total stellar mass of 500\,$M_{\odot}$ based on \cite{torii11}, the crossing time estimated is $\sim$2.7\,Myr. This is 2$\sigma$ larger than the median age of the accreting PMS stars. Future spectroscopy of cluster members, that can estimate the radial velocity spread of the region is essential to confirm the crossing time \citep{reviewadamo}.


Fig.\,\ref{moneyplot} shows the positions of the accreting PMS stars, YSOs, and cold cores detected by \cite{tapia} in the Trifid nebula. They are expected to represent an age sequence of oldest to youngest stars currently forming in M\,20. These positions are overlaid on the H$\alpha$ VPHAS+ image, the {\it Spitzer} 8$\mu$m image, and the {\it Herschel} 350$\mu$m respectively for the reader to relatively gauge the position of the H$\alpha$ nebulosity, and the PAH and IR emission. Also shown are the 2\,km\,s$^{-1}$ and cloud C contours taken from \cite{torii11}. The approximate location of the central cavity is also given, and corresponds to the complementary distribution region in \cite{torii17} between $18^h02^m15^s-19^s$, and $-23^{\degr}03^{\arcmin}-01^{\arcmin}$. The boundaries of the molecular clouds can be visualised by the respective cloud contours given in Fig.\,\ref{moneyplot}.

From Fig.\,\ref{moneyplot} the positions of the accreting PMS stars and YSOs are seen to correlate to the positions of the molecular clouds. The PMS stars and Class\,II YSOs are preferentially found along the boundaries of the molecular clouds, with none found within the boundaries of the central cavity. Class\,II YSOs are also located along the edges of the dust lanes. Further, the ages of the stars show no clear demarcation between old or young stars, besides a group of older stars lying outside the displayed boundaries of the molecular clouds to the south east. The absence of PMS stars near the centre is not decisive evidence, as the PMS sample is incomplete spatially. However, also no Class\,II YSOs are detected towards the centre of the Nebula. This can be considered as evidence as the {\it Spitzer} photometry is not affected towards the central region when compared to the optical photometry.

In contrast, the Class\,I YSOs are found favouring the dust lanes near the centre of the nebula. In addition, a concentration of Class\,I YSOs and cold cores are found towards the South West, outside clearly visible nebulosity in H$\alpha$, but near nebulous filaments visible in the 8$\mu$m image. This suggests that a newer generation of stars is being formed near the periphery, where the cold cores, and the Class I sources are located. A single flat source is detected towards the central region, and few Class I sources, and the cold core TC1 are found near HD\,164992A.

From the ages and positions of the PMS stars, YSOs, and cold core samples compared to the colliding molecular clouds in Fig.\,\ref{moneyplot}, there is evidence suggestive of CCC induced star formation. The locations of the stars are along the edges of the two clouds with similar ages. This age is smaller than the crossing timescale $\sim$2.7\,Myr. No stars are found near the centre of the two colliding clouds, suggestive of a cavity devoid of on-going star formation. The PMS stars, Class\,II YSOs have ages $\sim$1.5\,Myr. Note that a small grouping of Class I YSOs, and cold cores to the south west of the nebula, falling on the southern filament are younger than most of the stellar population in the region, and are forming in a subsequent outburst of star formation. 




In summary the gravitational energy of the clouds is not sufficient to bind them, and the heating from HD\,164492A seems insufficient to cause the observed expansion. The velocity differences between the two clouds is systematic. Travelling backwards, they must have collided $\sim$0.7\,Myr. The collision of the small cloud created a cavity on the larger cloud where the two clouds collide and now overlap in the projected space \citep{torii17}. Star formation was triggered likely immediately after, and molecular material across the two clouds collapsed to form the currently visible young stellar population roughly a million years ago. The positions and ages of YSOs and PMS stars lend credence to the hypothesis that they may have formed after CCC. The Class\,II YSOs and PMS stars ($\sim$1.5\,Myr) are located preferentially towards the edges of the clouds, and not found in the centre. Class\,I YSOs ($\lesssim0.5$\,Myr) are found more concentrated around the centre of the clouds. The stars on either ends of the molecular cloud are around the same age, which is less than the  crossing time of $\sim$2.7\,Myr. 

A tightness in the dynamical age (0.85\,Myr) and the age of the accreting PMS stars may be alleviated if there are precise reddening, and distance estimates. Further spectroscopic observations are needed to confirm this. Such spectroscopic observations can also potentially identify discless/non-accreting members. While the ages of such members are in keeping with the general age of the region they reside in, they are often spatially less concentrated (although occupying similar spatial distributions) compared to their more active brethren.

\section{Conclusions}

\begin{enumerate}
    \item We have identified 46 Class\,II, 9 Flat, and 6 Class\,I\,YSOs in the Trifid Nebula on the basis of their infrared SED slope, $\alpha$, and positions in the mIR colour-colour plane. No Class\,III YSOs are included in our analysis.
    \item 41 accreting PMS stars were identified on the basis of H$\alpha$ excess against $ri$H$\alpha$ photometry. The identified stars have proper motions indicating they likely belong to the Trifid Nebula. 90\% of the accreting PMS stars have nIR colours indicating the presence of a circumstellar dust disc. 
    \item The distance to the Trifid Nebula measured from {\it Gaia} EDR3 parallaxes of a subset of cluster members is around 1250\,pc, placing the region in the Sagittarius arm of the Milky Way.
    \item Based on the positions and ages of the young stellar population in the region, it is likely that star formation was triggered by the collision of two clouds $\sim$1\,Myr ago, leading to the currently observed distribution. 
\end{enumerate}

\begin{acknowledgments}
V.M.K. acknowledges funding from CONICYT Programa de Astronomia Fondo Gemini-Conicyt N$^{\rm o}$ 32RF180005. The work of V.M.K. is supported by NOIRLab, which is managed by the Association of Universities for Research in Astronomy (AURA) under a cooperative agreement with the National Science Foundation. V.M.K. thanks K. Torii for kindly providing the molecular cloud cubes, and the anonymous referee for detailed comments which helped improve this paper. This work is based in part on observations made with the Spitzer Space Telescope, which is operated by the Jet Propulsion Laboratory, California Institute of Technology under a contract with NASA. This work is based in part on data obtained as part of the UKIRT Infrared Deep Sky Survey. Based in part on observations made with ESO Telescopes at the La Silla or Paranal Observatories under programme ID(s) 177.D-3023(B), 177.D-3023(C), 177.D-3023(D), 177.D-3023(E). This work presents in part results from the European Space Agency (ESA) space mission Gaia. Gaia data are being processed by the Gaia Data Processing and Analysis Consortium (DPAC). Funding for the DPAC is provided by national institutions, in particular the institutions participating in the Gaia MultiLateral Agreement (MLA). Herschel is an ESA space observatory with science instruments provided by European-led Principal Investigator consortia and with important participation from NASA. 
\end{acknowledgments}

%

\vspace{5mm}
\facilities{{\it{Spitzer}},{\it{Herschel}}, VLT(VST), {\it{Gaia}}}


\software{APLpy \citep{aplpy}, FILFinder \citep{koch15}}

\bibliography{sample631}{}

\begin{thebibliography}{}
\expandafter\ifx\csname natexlab\endcsname\relax\def\natexlab#1{#1}\fi
\providecommand{\url}[1]{\href{#1}{#1}}
\providecommand{\dodoi}[1]{doi:~\href{http://doi.org/#1}{\nolinkurl{#1}}}
\providecommand{\doeprint}[1]{\href{http://ascl.net/#1}{\nolinkurl{http://ascl.net/#1}}}
\providecommand{\doarXiv}[1]{\href{https://arxiv.org/abs/#1}{\nolinkurl{https://arxiv.org/abs/#1}}}

\bibitem[{{Adamo} {et~al.}(2020){Adamo}, {Zeidler}, {Kruijssen}, {Chevance},
  {Gieles}, {Calzetti}, {Charbonnel}, {Zinnecker}, \& {Krause}}]{reviewadamo}
{Adamo}, A., {Zeidler}, P., {Kruijssen}, J.~M.~D., {et~al.} 2020, \ssr, 216,
  69, \dodoi{10.1007/s11214-020-00690-x}

\bibitem[{{Allen} {et~al.}(2004){Allen}, {Calvet}, {D'Alessio}, {Merin},
  {Hartmann}, {Megeath}, {Gutermuth}, {Muzerolle}, {Pipher}, {Myers}, \&
  {Fazio}}]{allen}
{Allen}, L.~E., {Calvet}, N., {D'Alessio}, P., {et~al.} 2004, \apjs, 154, 363,
  \dodoi{10.1086/422715}

\bibitem[{{Bailer-Jones} {et~al.}(2018){Bailer-Jones}, {Rybizki}, {Fouesneau},
  {Mantelet}, \& {Andrae}}]{bailerjones}
{Bailer-Jones}, C.~A.~L., {Rybizki}, J., {Fouesneau}, M., {Mantelet}, G., \&
  {Andrae}, R. 2018, \aj, 156, 58, \dodoi{10.3847/1538-3881/aacb21}

\bibitem[{{Barrado y Navascu{\'e}s} \& {Mart{\'{\i}}n}(2003)}]{barr03}
{Barrado y Navascu{\'e}s}, D., \& {Mart{\'{\i}}n}, E.~L. 2003, AJ, 126, 2997,
  \dodoi{10.1086/379673}

\bibitem[{{Bressan} {et~al.}(2012){Bressan}, {Marigo}, {Girardi}, {Salasnich},
  {Dal Cero}, {Rubele}, \& {Nanni}}]{bress12}
{Bressan}, A., {Marigo}, P., {Girardi}, L., {et~al.} 2012, MNRAS, 427, 127,
  \dodoi{10.1111/j.1365-2966.2012.21948.x}

\bibitem[{{Cambr{\'e}sy} {et~al.}(2011){Cambr{\'e}sy}, {Rho}, {Marshall}, \&
  {Reach}}]{cambresy11}
{Cambr{\'e}sy}, L., {Rho}, J., {Marshall}, D.~J., \& {Reach}, W.~T. 2011, \aap,
  527, A141, \dodoi{10.1051/0004-6361/201015863}

\bibitem[{{Cernicharo} {et~al.}(1998){Cernicharo}, {Lefloch}, {Cox},
  {Cesarsky}, {Esteban}, {Yusef-Zadeh}, {Mendez}, {Acosta-Pulido}, {Garcia
  Lopez}, \& {Heras}}]{Cernicharo}
{Cernicharo}, J., {Lefloch}, B., {Cox}, P., {et~al.} 1998, Science, 282, 462,
  \dodoi{10.1126/science.282.5388.462}

\bibitem[{{Churchwell} {et~al.}(2009){Churchwell}, {Babler}, {Meade},
  {Whitney}, {Benjamin}, {Indebetouw}, {Cyganowski}, {Robitaille}, {Povich},
  {Watson}, \& {Bracker}}]{glimpse}
{Churchwell}, E., {Babler}, B.~L., {Meade}, M.~R., {et~al.} 2009, \pasp, 121,
  213, \dodoi{10.1086/597811}

\bibitem[{{Cutri} {et~al.}(2003){Cutri}, {Skrutskie}, {van Dyk}, {Beichman},
  {Carpenter}, {Chester}, {Cambresy}, \& et~al.}]{cutri03}
{Cutri}, R.~M., {Skrutskie}, M.~F., {van Dyk}, S., {et~al.} 2003, {2MASS All
  Sky Catalog of point sources.}

\bibitem[{{D'Alessio} {et~al.}(2001){D'Alessio}, {Calvet}, \& {Hartmann}}]{d05}
{D'Alessio}, P., {Calvet}, N., \& {Hartmann}, L. 2001, ApJ, 553, 321,
  \dodoi{10.1086/320655}

\bibitem[{{De Marchi} {et~al.}(2010){De Marchi}, {Panagia}, \&
  {Romaniello}}]{de10}
{De Marchi}, G., {Panagia}, N., \& {Romaniello}, M. 2010, ApJ, 715, 1,
  \dodoi{10.1088/0004-637X/715/1/1}

\bibitem[{{Dotter} {et~al.}(2008){Dotter}, {Chaboyer}, {Jevremovi{\'c}},
  {Kostov}, {Baron}, \& {Ferguson}}]{dotter}
{Dotter}, A., {Chaboyer}, B., {Jevremovi{\'c}}, D., {et~al.} 2008, \apjs, 178,
  89, \dodoi{10.1086/589654}

\bibitem[{{Drew} {et~al.}(2014){Drew}, {Gonzalez-Solares}, {Greimel}, {Irwin},
  {K{\"u}pc{\"u} Yoldas}, {Lewis}, {Barentsen}, {Eisl{\"o}ffel}, \&
  et~al..}]{drew14}
{Drew}, J.~E., {Gonzalez-Solares}, E., {Greimel}, R., {et~al.} 2014, MNRAS,
  440, 2036, \dodoi{10.1093/mnras/stu394}

\bibitem[{{Feigelson} {et~al.}(2013){Feigelson}, {Townsley}, {Broos}, {Busk},
  {Getman}, {King}, {Kuhn}, {Naylor}, {Povich}, {Baddeley}, {Bate},
  {Indebetouw}, {Luhman}, {McCaughrean}, {Pittard}, {Pudritz}, {Sills}, {Song},
  \& {Wadsley}}]{mystix}
{Feigelson}, E.~D., {Townsley}, L.~K., {Broos}, P.~S., {et~al.} 2013, \apjs,
  209, 26, \dodoi{10.1088/0067-0049/209/2/26}

\bibitem[{{Fukui} {et~al.}(2020){Fukui}, {Habe}, {Inoue}, {Enokiya}, \&
  {Tachihara}}]{fukui}
{Fukui}, Y., {Habe}, A., {Inoue}, T., {Enokiya}, R., \& {Tachihara}, K. 2020,
  \pasj, \dodoi{10.1093/pasj/psaa103}

\bibitem[{{Gaia Collaboration} {et~al.}(2020){Gaia Collaboration}, {Brown},
  {Vallenari}, {Prusti}, {de Bruijne}, {Babusiaux}, \& {Biermann}}]{gaia3}
{Gaia Collaboration}, {Brown}, A.~G.~A., {Vallenari}, A., {et~al.} 2020, arXiv
  e-prints, arXiv:2012.01533.
\newblock \doarXiv{2012.01533}

\bibitem[{{Gullbring} {et~al.}(1998){Gullbring}, {Hartmann}, {Briceno}, \&
  {Calvet}}]{gullbring98}
{Gullbring}, E., {Hartmann}, L., {Briceno}, C., \& {Calvet}, N. 1998, ApJ, 492,
  323, \dodoi{10.1086/305032}

\bibitem[{{Gutermuth} {et~al.}(2009){Gutermuth}, {Megeath}, {Myers}, {Allen},
  {Pipher}, \& {Fazio}}]{gut}
{Gutermuth}, R.~A., {Megeath}, S.~T., {Myers}, P.~C., {et~al.} 2009, \apjs,
  184, 18, \dodoi{10.1088/0067-0049/184/1/18}

\bibitem[{{Harvey} {et~al.}(2006){Harvey}, {Chapman}, {Lai}, \&
  {Evans}}]{harvey}
{Harvey}, P.~M., {Chapman}, N., {Lai}, S.-P., \& {Evans}, Neal~J., I. e.~a.
  2006, \apj, 644, 307, \dodoi{10.1086/503520}

\bibitem[{{Indebetouw} {et~al.}(2005){Indebetouw}, {Mathis}, {Babler}, {Meade},
  {Watson}, {Whitney}, {Wolff}, {Wolfire}, {Cohen}, {Bania}, {Benjamin},
  {Clemens}, {Dickey}, {Jackson}, {Kobulnicky}, {Marston}, {Mercer},
  {Stauffer}, {Stolovy}, \& {Churchwell}}]{inde05}
{Indebetouw}, R., {Mathis}, J.~S., {Babler}, B.~L., {et~al.} 2005, \apj, 619,
  931, \dodoi{10.1086/426679}

\bibitem[{{Kalari}(2019)}]{kalari19}
{Kalari}, V.~M. 2019, \mnras, 484, 5102, \dodoi{10.1093/mnras/stz250}

\bibitem[{{Kalari} {et~al.}(2015){Kalari}, {Vink}, {Drew}, {Barentsen},
  {Drake}, {Eisl{\"o}ffel}, {Mart{\'{\i}}n}, {Parker}, {Unruh}, {Walton}, \&
  {Wright}}]{Kalari15}
{Kalari}, V.~M., {Vink}, J.~S., {Drew}, J.~E., {et~al.} 2015, MNRAS, 453, 1026.
\newblock \doarXiv{1507.06786}

\bibitem[{{Koch} \& {Rosolowsky}(2015)}]{koch15}
{Koch}, E.~W., \& {Rosolowsky}, E.~W. 2015, \mnras, 452, 3435,
  \dodoi{10.1093/mnras/stv1521}

\bibitem[{{Lefloch} {et~al.}(2002){Lefloch}, {Cernicharo}, {Rodr{\'\i}guez},
  {Miville-Desch{\^e}nes}, {Cesarsky}, \& {Heras}}]{lefloch}
{Lefloch}, B., {Cernicharo}, J., {Rodr{\'\i}guez}, L.~F., {et~al.} 2002, \apj,
  581, 335, \dodoi{10.1086/344049}

\bibitem[{{Loren}(1976)}]{loren1976}
{Loren}, R.~B. 1976, \apj, 209, 466, \dodoi{10.1086/154741}

\bibitem[{{Lucas} {et~al.}(2008){Lucas}, {Hoare}, {Longmore}, {Schr{\"o}der},
  {Davis}, {Adamson}, {Bandyopadhyay}, {de Grijs}, \& {Smith}}]{ukidss}
{Lucas}, P.~W., {Hoare}, M.~G., {Longmore}, A., {et~al.} 2008, MNRAS, 391, 136,
  \dodoi{10.1111/j.1365-2966.2008.13924.x}

\bibitem[{{McKee} \& {Ostriker}(2007)}]{mckee}
{McKee}, C.~F., \& {Ostriker}, E.~C. 2007, \araa, 45, 565,
  \dodoi{10.1146/annurev.astro.45.051806.110602}

\bibitem[{{Megeath} {et~al.}(2004){Megeath}, {Allen}, {Gutermuth}, {Pipher},
  {Myers}, {Calvet}, {Hartmann}, {Muzerolle}, \& {Fazio}}]{megeath}
{Megeath}, S.~T., {Allen}, L.~E., {Gutermuth}, R.~A., {et~al.} 2004, \apjs,
  154, 367, \dodoi{10.1086/422823}

\bibitem[{{Meyer} {et~al.}(1997){Meyer}, {Calvet}, \& {Hillenbrand}}]{meyer97}
{Meyer}, M.~R., {Calvet}, N., \& {Hillenbrand}, L.~A. 1997, AJ, 114, 288,
  \dodoi{10.1086/118474}

\bibitem[{{Pecaut} \& {Mamajek}(2013)}]{mamajek}
{Pecaut}, M.~J., \& {Mamajek}, E.~E. 2013, \apjs, 208, 9,
  \dodoi{10.1088/0067-0049/208/1/9}

\bibitem[{{Petit} {et~al.}(2019){Petit}, {Wade}, {Schneider}, {Fossati},
  {Kamp}, {Neiner}, {David-Uraz}, {Alecian}, \& {MiMeS
  Collaboration}}]{164492Age}
{Petit}, V., {Wade}, G.~A., {Schneider}, F.~R.~N., {et~al.} 2019, \mnras, 489,
  5669, \dodoi{10.1093/mnras/stz2469}

\bibitem[{{Pickles}(1998)}]{pick98}
{Pickles}, A.~J. 1998, PASP, 110, 863, \dodoi{10.1086/316197}

\bibitem[{{Rho} {et~al.}(2001){Rho}, {Corcoran}, {Chu}, \& {Reach}}]{rho01}
{Rho}, J., {Corcoran}, M.~F., {Chu}, Y.-H., \& {Reach}, W.~T. 2001, \apj, 562,
  446, \dodoi{10.1086/323053}

\bibitem[{{Rho} {et~al.}(2008){Rho}, {Lefloch}, {Reach}, \&
  {Cernicharo}}]{rho08}
{Rho}, J., {Lefloch}, B., {Reach}, W.~T., \& {Cernicharo}, J. 2008, {M20: Star
  Formation in a Young HII Region}, ed. B.~{Reipurth}, Vol.~5, 509

\bibitem[{{Rho} {et~al.}(2006){Rho}, {Reach}, {Lefloch}, \& {Fazio}}]{rho06}
{Rho}, J., {Reach}, W.~T., {Lefloch}, B., \& {Fazio}, G.~G. 2006, \apj, 643,
  965, \dodoi{10.1086/503245}

\bibitem[{{Robitaille} \& {Bressert}(2012)}]{aplpy}
{Robitaille}, T., \& {Bressert}, E. 2012, {APLpy: Astronomical Plotting Library
  in Python}.
\newblock \doeprint{1208.017}

\bibitem[{{Robitaille} {et~al.}(2006){Robitaille}, {Whitney}, {Indebetouw},
  {Wood}, \& {Denzmore}}]{rob}
{Robitaille}, T.~P., {Whitney}, B.~A., {Indebetouw}, R., {Wood}, K., \&
  {Denzmore}, P. 2006, \apjs, 167, 256, \dodoi{10.1086/508424}

\bibitem[{{Siess} {et~al.}(2000){Siess}, {Dufour}, \& {Forestini}}]{siess00}
{Siess}, L., {Dufour}, E., \& {Forestini}, M. 2000, A\&A, 358, 593

\bibitem[{{Smith} {et~al.}(2002){Smith}, {Norris}, \& {Crowther}}]{smith02}
{Smith}, L.~J., {Norris}, R. P.~F., \& {Crowther}, P.~A. 2002, \mnras, 337,
  1309, \dodoi{10.1046/j.1365-8711.2002.06042.x}

\bibitem[{{Tapia} {et~al.}(2018){Tapia}, {Persi}, {Rom{\'a}n-Z{\'u}{\~n}iga},
  {Elia}, {Giovannelli}, \& {Sabau-Graziati}}]{tapia}
{Tapia}, M., {Persi}, P., {Rom{\'a}n-Z{\'u}{\~n}iga}, C., {et~al.} 2018,
  \mnras, 475, 3029, \dodoi{10.1093/mnras/sty048}

\bibitem[{{Torii} {et~al.}(2011){Torii}, {Enokiya}, {Sano}, {Yoshiike},
  {Hanaoka}, {Ohama}, {Furukawa}, {Dawson}, {Moribe}, {Oishi}, {Nakashima},
  {Okuda}, {Yamamoto}, {Kawamura}, {Mizuno}, {Maezawa}, {Onishi}, {Mizuno}, \&
  {Fukui}}]{torii11}
{Torii}, K., {Enokiya}, R., {Sano}, H., {et~al.} 2011, \apj, 738, 46,
  \dodoi{10.1088/0004-637X/738/1/46}

\bibitem[{{Torii} {et~al.}(2017){Torii}, {Hattori}, {Hasegawa}, {Ohama},
  {Haworth}, {Shima}, {Habe}, {Tachihara}, {Mizuno}, {Onishi}, {Mizuno}, \&
  {Fukui}}]{torii17}
{Torii}, K., {Hattori}, Y., {Hasegawa}, K., {et~al.} 2017, \apj, 835, 142,
  \dodoi{10.3847/1538-4357/835/2/142}

\bibitem[{{Tremblin} {et~al.}(2014){Tremblin}, {Anderson}, {Didelon}, {Raga},
  {Minier}, {Ntormousi}, {Pettitt}, {Pinto}, {Samal}, {Schneider}, \&
  {Zavagno}}]{tremblin}
{Tremblin}, P., {Anderson}, L.~D., {Didelon}, P., {et~al.} 2014, \aap, 568, A4,
  \dodoi{10.1051/0004-6361/201423959}

\bibitem[{{Urquhart} {et~al.}(2014){Urquhart}, {Csengeri}, {Wyrowski},
  {Schuller}, {Bontemps}, {Bronfman}, {Menten}, {Walmsley}, {Contreras},
  {Beuther}, {Wienen}, \& {Linz}}]{atlasgal}
{Urquhart}, J.~S., {Csengeri}, T., {Wyrowski}, F., {et~al.} 2014, \aap, 568,
  A41, \dodoi{10.1051/0004-6361/201424126}

\bibitem[{{Wade} {et~al.}(2017){Wade}, {Shultz}, {Sikora}, {Bernier},
  {Rivinius}, {Alecian}, {Petit}, {Grunhut}, \& {BinaMIcS
  Collaboration}}]{wade}
{Wade}, G.~A., {Shultz}, M., {Sikora}, J., {et~al.} 2017, \mnras, 465, 2517,
  \dodoi{10.1093/mnras/stw2799}

\bibitem[{{Yusef-Zadeh} {et~al.}(2005){Yusef-Zadeh}, {Biretta}, \&
  {Geballe}}]{yusefzadeh05}
{Yusef-Zadeh}, F., {Biretta}, J., \& {Geballe}, T.~R. 2005, \aj, 130, 1171,
  \dodoi{10.1086/432095}

\bibitem[{{Yusef-Zadeh} {et~al.}(2000){Yusef-Zadeh}, {Shure}, {Wardle}, \&
  {Kassim}}]{yusefzadeh00}
{Yusef-Zadeh}, F., {Shure}, M., {Wardle}, M., \& {Kassim}, N. 2000, \apj, 540,
  842, \dodoi{10.1086/309352}

\end{thebibliography}
\bibliographystyle{aasjournal}



\end{document}